\documentclass[fleqn,usenatbib]{mnras}

\usepackage{newtxtext,newtxmath}

\usepackage[T1]{fontenc}
\usepackage{ae,aecompl}

\usepackage{graphicx}	
\usepackage{amsmath}	
\usepackage{pdflscape}
\usepackage{multirow}

\title[Chemical evolution of CALIFA galaxies]{Evolution of the chemical enrichment and the Mass-Metallicity relation in CALIFA galaxies}

\author[A. Camps-Fari\~{n}a et al.]{A. Camps-Fari\~{n}a$^{1}$\thanks{E-mail:acamps@astro.unam.mx}, S. F. Sanchez$^{1}$, E. A. D. Lacerda$^{1}$, L. Carigi$^{1}$,\newauthor R. Garc\'{i}a-Benito$^{2}$, D. Mast$^{3,4}$, L. Galbany$^{5}$\\
$^{1}$Instituto de Astronom\'ia, Universidad Nacional Aut\'onoma de M\'exico, Apartado Postal 70-264, CP 04510 Ciudad de M\'exico, M\'exico\\
$^{2}$Instituto de Astrofísica de Andalucía (CSIC), PO Box 3004, 18080 Granada, Spain\\
$^{3}$Observatorio Astron\'{o}mico de C\'{o}rdoba, Laprida 854, X5000BGR, C\'{o}rdoba, Argentina.\\
$^{4}$Consejo de Investigaciones Cient\'{i}ficas y T\'ecnicas de la Rep\'ublica Argentina, Avda. Rivadavia 1917, C1033AAJ, CABA, Argentina.\\
$^{5}$Departamento de F\'isica Te\'orica y del Cosmos, Universidad de Granada, E-18071 Granada, Spain.
}

\begin{document}

\pagerange{\pageref{firstpage}--\pageref{lastpage}} \pubyear{2002}
\maketitle
\label{firstpage}

\begin{abstract}
We apply fossil record techniques to the CALIFA sample to study how galaxies in the local universe have evolved in terms of their chemical content. We show how the stellar metallicity and the mass-metallicity relation (MZR) evolve through time for the galaxies in our sample and how this evolution varies when we divide them based on their mass, morphology and star-forming status. We also check the impact of measuring the metallicity at the centre or the outskirts. We find the expected results that the most massive galaxies were enriched more quickly, and that the MZR was steeper at higher redshifts. However, once we separate the galaxies into morphology bins this behaviour is less clear, which suggests that morphology is a primary factor in determining how quickly a galaxy becomes enriched, but with mass determining the final enrichment. We also find that star-forming galaxies appear to be asymptotic in their chemical evolution, that is, the metallicity of star-forming galaxies of any mass is very similar at recent times unlike several Gyr ago.

\end{abstract}


\section{Introduction}
The interstellar medium (ISM) and stars form a cycle in which the latter are born from the former. The ISM itself has been enriched with metals from the previous generations of stars. Finally, these new stars will in turn further enrich the ISM in which the following generations of stars are formed, thus closing the duty-cycle. The metal content of the ISM at a particular time is effectively "locked" in the chemical composition of the stars that were formed at that time.

As galaxies evolve, their metal content generally increases as a result of star formation \citep{Tinsley1980, McWilliam1997}, while other physical processes, such as outflows and inflows, are capable of removing or diluting it (\citealt{Sancisi2008, Dave2011, Lilly2013, Belfiore2016, Barrera-Ballesteros2018}). These processes, as well as star formation itself, depend on the physical parameters of the galaxy such as the halo, stellar, and gas masses or the morphology and as a result the metal content of galaxies also correlates with them. \cite{Lequeux1979} pioneered such studies showing that the total mass of a galaxy, inferred from the rotational velocity, correlates with the metallicity. \cite{Tremonti2004} used 53,000 galaxies from the SDSS survey to show that the physical parameter that gave the tightest correlation with the metallicity is the mass of the galaxy. Further explorations have confirmed that relation using non aperture biased spatially-resolved or integrated spectroscopic data (e.g. \citealt{Rosales-Ortega2012, Sanchez2013, Zahid2014, Lara-Lopez2013, Sanchez2017}).

The reason the mass-metallicity relation (MZR) is a fundamental relation for the evolution of galaxies is that its shape is directly related to the processes that regulate the star-formation in galaxies. At low masses the relation is nearly linear, which is expected of a closed-box model. At high masses, though, the MZR flattens towards an asymptotic value of the metallicity \citep{Tremonti2004}. This value depends on the calibrator used \citep{Kewley2008, Barrera-Ballesteros2017, Sanchez2017, Sanchez2019b} as well as on the shape of the IMF which affects metal production with a steeper IMF producing less metals \citep{Lian2018}.

The MZR is usually calculated based on global measurements of galaxies, but there is a corresponding relation inside a galaxy: by comparing the surface mass to the local metallicity we obtain the resolved MZR or rMZR. This relation and how it compares to its global counterpart are ongoing topics of research \citep{Moran2012, Rosales-Ortega2012, Sanchez2013, Erroz-Ferrer2019}, but in this article we are interested only in the properties of the global MZR.

The studies mentioned so far use the nebular metallicity rather than the metallicity of the stars. Nebular metallicity is measured from the properties of the emission lines produced by the ionised gas. The oxygen abundance is a well-established proxy for this quantity and it is much easier to measure in HII regions than its abundance in stars owing to its intense interstellar emission lines this element produces. In order to obtain the metallicity for the stars the stellar continuum needs to be fitted with stellar templates, using their shapes and absorption lines to recover the properties of the stellar population which produce weaker features on a spectrum.

However, the nebular metallicity can be measured only in gas-rich, star-forming galaxies and it gives us information regarding only the current values of the metallicity in a galaxy. The evolution of the gas metallicity can be probed by observing galaxies at increasing redshift. Studies have generally found that galaxies have a growth in metallicity on cosmological timescales (e.g. \citealt{Steidel2014, Troncoso2014, Wuyts2014, Sanders2015, Onodera2015, Kashino2017, Cullen2019}). The speed of this growth depends on mass in agreement with the observed phenomenon of downsizing which shows that the more massive galaxies appear to have evolved more quickly in all aspects, assembling their masses earlier and also becoming enriched more quickly. \cite{Spitoni2020} show how imposing that galaxies follow both the nebular MZR and the SFMS into an evolutionary model causes downsizing to appear, showing the intrinsic ties between downsizing and these scaling relations.

Stellar metallicity is explored using a completely different technique: the fossil record method. This technique allows us to fit templates associated with stellar populations of a particular age and metallicity (or having a particular star-formation history, e.g., \citealt{Bitsakis2018}) to the observed stellar spectra or particular features in that spectra, such as stellar indices \citep{Kauffmann2003}. We can use the recovered weights for each population to analyse the current stellar composition of the galaxy or, as is the goal of this article, to estimate the evolution of the metallicity and the mass-metallicity relation in galaxies. A particularly noteworthy advantage of this type of analysis is that we can measure the evolution of the same sample of galaxies through time, rather than observing different galaxies at different times as is the case for high-redshift observations.

Pioneering work on this topic was produced by \cite{Panter2008}, who determined the MZR of SDSS galaxies at different redshifts using simple stellar population (SSP) fitting and then compared those results to the ones from \cite{Tremonti2004} and similar studies which measured gas metallicities. 
\cite{Vale-Asari2009} measured stellar metallicities of SDSS galaxies derived from fitting SSP templates with \textsc{starlight} \citep{CidFernandes2005} to show the evolution of the stellar MZR slope with time. Previously, \cite{Gallazzi2005} proved that the stellar MZR presents a shape similar to that reported for the nebular MZR. This shape seems to be preserved in the spatially resolved MZR with both of them showing a clear evolution with enrichment for massive galaxies \citep{Gonzalez-Delgado2014b} 

In this article we use the CALIFA dataset of Integral Field Spectroscopy (IFS) observations of galaxies in the nearby universe to measure the metallicity and mass-metallicity relation at different epochs during the galaxies' lifetimes. We also separate the galaxies according to their mass (for the evolution of the metallicity), morphology and star forming status and analysed the impact of these on the results. The metallicities were measured at the effective radius, but also checked how the results vary with galactocentric distance.

The article is organised as follows: in Section 2 we describe the sample of galaxies that we use and their data; in Section 3 we describe the analysis of the data with a summary of the products of the reduction and analysis pipeline for the survey, including a description of how the quantities were calculated and the method used to average the quantities in a consistent manner; in Section 4 we present the results of the analysis, showing the evolution of the metallicity and the mass-metallicity relation, as well as a proxy for the star-formation rate. We also show how the evolution of these parameters is affected by the galaxies' morphology and star-forming status, as well as the galactocentric distance at which they are measured. Finally, in Section 5 we discuss the validity of our results and the impact and relevance that they have on our understanding of the evolution of galaxies.

\section{Sample and Data}
\label{sec:sample}
The base sample we used consists of all galaxies from the CALIFA survey with good spectroscopic data up to DR3 using the low-resolution setup (V500). In addition, we included objects observed by CALIFA extended surveys \citep{Sanchez2016}, including those of the PMAS/Ppak Integral-field Supernova hosts COmpilation \citep[PISCO; ][]{Galbany2018}. This results in an initial sample of 877 galaxies. This sample follows the criteria of the CALIFA mother sample \citep[MS; ][]{Walcher2014} so that all objects have their optical extension inside the FoV of the instrument, but some could be slightly fainter or brighter than those in the original MS. Moreover, some objects have their redshifts slightly outside the boundary of the CALIFA original selection (with this extended sample $z$ goes up to 0.08, but with 93 per cent of the galaxies below 0.035). We further refined the sample by removing those galaxies which are known to host an AGN \citep{Lacerda2020} as well as those with an inclination above 70\degr{} or which appear to be interacting. After this selection we ended with the final sample of 529 galaxies which we use in this article, comprising a stellar-mass range of $10^{7.8} - 10^{12.2}$ M$_\odot$ with the mean value of $10^{10.4}$ M$_\odot$.

The CALIFA V500 setup adopted within this study covers a wavelength range between 3745\AA\ and 7200\AA, providing a spectral resolution with FWHM$\sim$6.5\AA, throughout the considered spectral range. The PPAK fibre bundle \citep{Kelz2006} has a central bundle of 331 fibres of 2.7$\arcsec$/diameter, that cover an hexagonal field-of-view of 74$\arcsec$ $\times$64$\arcsec$ with a filling factor of slightly less than a 60\% (an additional set of 36 fibres are distributed in 6 mini-bundles far away from this central bundle to sample the night-sky). In order to obtain a 100\% coverage of the FoV, while also increasing slightly the spatial resolution, a three-pointing dithering scheme was adopted during the observations considered here \citep{Sanchez2012}. Thus, after a standard reduction procedure \citep{Sanchez2006} the final product is a data-cube with two dimensions corresponding to the spatial coordinates and the third one sampling the wavelength range. The reduction comprises spectral extraction, wavelength calibration, sky subtraction, fibre transmission correction, flux calibration, cube reconstruction, differential atmospheric correction, and image registration. The variance was propagated through the different reduction steps and stored in an additional data-cubes, together with a mask of the bad pixels (including cosmic rays and vignetting), and a map of the variance covariance due to the different reduction steps. The final data-cubes have a spatial resolution of FWHM$\sim$2.5$\arcsec$ and a sampling of 1$\arcsec$, and a photometric accuracy better than a 5\% through all the considered wavelength range. For further details on the data, the reduction and the quality, we refer the reader to \cite{Husemann2013, Garcia-Benito2015, Sanchez2016}. 

The morphological classification for the galaxies from the DR3 sample is obtained from \cite{Walcher2014}. The classification of the remaining galaxies is conducted using the same scheme employed by \cite{Walcher2014}. The procedure involves a by-eye inspection of the true-colour SDSS \citep[DR7; ][]{Abazajian2009} images. For those galaxies that are not present in this set of images, we use (i) similar images extracted from the CALIFA datacubes, which have lower spatial resolution, and (ii) emission-lines maps (described later on), in order to trace possible spiral arms structures. Finally, the morphological distribution is: 131 ellipticals (E0-E7), 73 lenticulars (S0+S0a), 325 spirals (Sa-Sm) and irregular (I) galaxies. This classification scheme was adopted previously in \cite{Lacerda2020} and \cite{Espinosa-Ponce2020} for the same sample of galaxies.


\section{Analysis}

\subsection{Pipe3D}

The stellar population synthesis of all datacubes is performed by {\em Pipe3D} \citep{Sanchez2016a,Sanchez2016b}, adopting the GSD156 library of SSP \citep{CidFernandes2013}. The GSD base is a combination of the SSP spectra provided by \cite{Vazdekis2010} for populations older than 63 Myr and the \cite{Gonzalez-Delgado2005} models for younger ages. This SSP-library includes 156 different templates comprising 39 stellar ages (from 0.001 to 14.1 Gyr, with uneven steps) and four metallicities (0.2, 0.4, 1 and 1.5 Z/Z$_\odot$) using the initial mass function (IMF) by \citep{Salpeter1955}. A wide set of previous studies has used this same library \citep{Perez2013, Gonzalez-Delgado2014b,Cano-Diaz2016,Sanchez-Menguiano2016, Ibarra-Medel2016,Sanchez2018,Lacerda2020}, where one can find more details and caveats about the fitting procedure.

The basic result of the stellar population analysis performed by {\rm Pipe3D} is the
fraction of light in a certain wavelength (the V-band in our case) that each SSP contributes
to the observed spectrum. This fraction of light (or weight) can be transformed to luminosity (considering the monochromatic luminosity at the considered fraction), and finally to a stellar mass (M$_*$) via the Mass-to-Light ratio of each SSP. Thus, the analysis provides with the amount of mass (and light) corresponding to stars of a certain age and metallicity. The first moments of this distribution over the range of ages and metallicities are the so-called Mass- (Luminosity-) weighted age and metallicity (LW and MW, respectively, e.g., \citealt{Sanchez2016}). These mean values are frequently used to characterise the stellar population, despite the fact that they are very rough estimation of the real distribution of ages (and metallicities).

By correcting the described mass distributions for the M$_*$ reduction as a result of star deaths that must have occurred from a particular time to the present (i.e., the look-back time corresponding to the age of the stars corrected by the redshift of the galaxy) it is possible to derive the amount of M$_*$ formed at any previous time (discretized by the age sampling of the SSP). The corresponding cumulative function from the beginning of the Universe to the current time is known as the Mass-Assembly History (MAH). The MAH, and its derivative with respect to time (the Star-formation History, SFH) has been extensively used to determine which galaxies (and regions of galaxies) form earlier and at what rate, and is the key evidence for galaxy downsizing: massive galaxies assemble their mass earlier and faster than less massive ones \citep{Thomas2005, Thomas2010, Perez-Gonzalez2008}. It has been used too to demonstrate the so-called local/resolved downsizing \citep{Perez2013, Ibarra-Medel2016, Garcia-Benito2017}: more massive regions in galaxies (ie., the centre) are formed before and more quickly than less massive regions (i.e. the outer areas).

The fraction of light provided by this analysis can be used to derive not only the stellar-mass distribution as a function of time, but also a similar distribution for the stellar metallicity. In this case, at any look-back time, it is possible to derive the LW- (and MW-) metallicity of the stars formed before that time. They are estimated as the first moment of the metallicity distribution for all ages older than the considered look-back time (LBT), correcting the fractions for the mass-loss from their formation to the observing time (i.e., the considered LBT). Estimated during the full Hubble time they correspond to the LW and MW- stellar metallicities described before. The metallicity distribution at each LBT comprises the chemical enrichment history (ChEH) of a galaxy (or a region of a galaxy). 

We should note here that this derivation is based on the assumption that the decomposition of the stellar population into the adopted SSP is accurate, precise, and unique. The true derivation differs considerably from this assumption. Different authors have already explored the many degeneracies inherent in the adopted analysis. We point the reader to \cite{Walcher2011, Sanchez-Blazquez2011} to understand the limitations of the adopted methodology. During this study we have intentionally ignored those limitations, so the current results are based on a single adopted methodology, a particular SSP library, and a considered dataset.

\subsection{Averaging properties}
\label{sec:averages}

In order to obtain average properties of a set of galaxies we need to make sure that the measurements are made on consistent physical scales so that the results are directly comparable. This allows us to stack the properties and perform averages for each bin. For the spatial scale we use the effective radius (Re; defined as the radius which contains half the light emitted by the galaxy) as a scaling quantity to average the properties of galaxies of different sizes. The metallicities for each galaxy are calculated at Re, which is a good indicator of the global metallicity of a galaxy \citep{Sanchez2017}. Many other physical quantities measured at the Re seem to be good proxies of the average properties of galaxies \citep{Gonzalez-Delgado2015, Sanchez2020}. Pipe3D also provides the values for the slope of the metallicity gradient, which will allow us to check the effects of measuring the metallicity at Re instead of at the centre or the outskirts (defined here as twice the Re).

The time scale of the data, on the other hand, needs to be corrected. While all the galaxies in the sample are located within the nearby universe they still have different redshifts. This means that we are effectively observing each galaxy at a slightly different cosmic time due to a difference in how long it took the light we are observing now to arrive. This effect is accounted for in our data by correcting the age of the fitted stellar templates with the redshift. The individual ChEHs, then, are shifted from one another in time. This affects the edges of the timeline the most: the more recent and more remote LBT in the timeline are not populated by all galaxies. Thus, we re-sample the individual time-evolution tracks of the metallicity to a common single timeline corresponding to the median of the individual time-lines, corresponding to the median redshift. A similar correction (and effect) was described in the analysis of the Mass-Assembly histories described by \cite{Ibarra-Medel2016}. However, in their case the effect is stronger due to a much wider range in redshift of the bulk population of analysed galaxies (extracted from the MaNGA survey, \citealt{Bundy2015}, that reaches up to $z\sim$0.17).

Once we have determined the representative timeline we re-sample each of the ChEHs using linear interpolation, discarding those values that lie out of the range. This method does not guarantee that all the LBT points contain all the galaxies in the sample, as indicated before. If we wanted to ensure this we would have needed to choose the time-line as containing only the overlap between all galaxies in the sample, discarding data points corresponding to LBT values which are covered by a sufficient number of galaxies to be statistically significant but which do not contain all of them. We chose the median time-line as a good compromise to obtain a good range of LBT and avoid a lack of statistical significance at the edges.

\begin{figure}
\centering
\includegraphics[width=\linewidth]{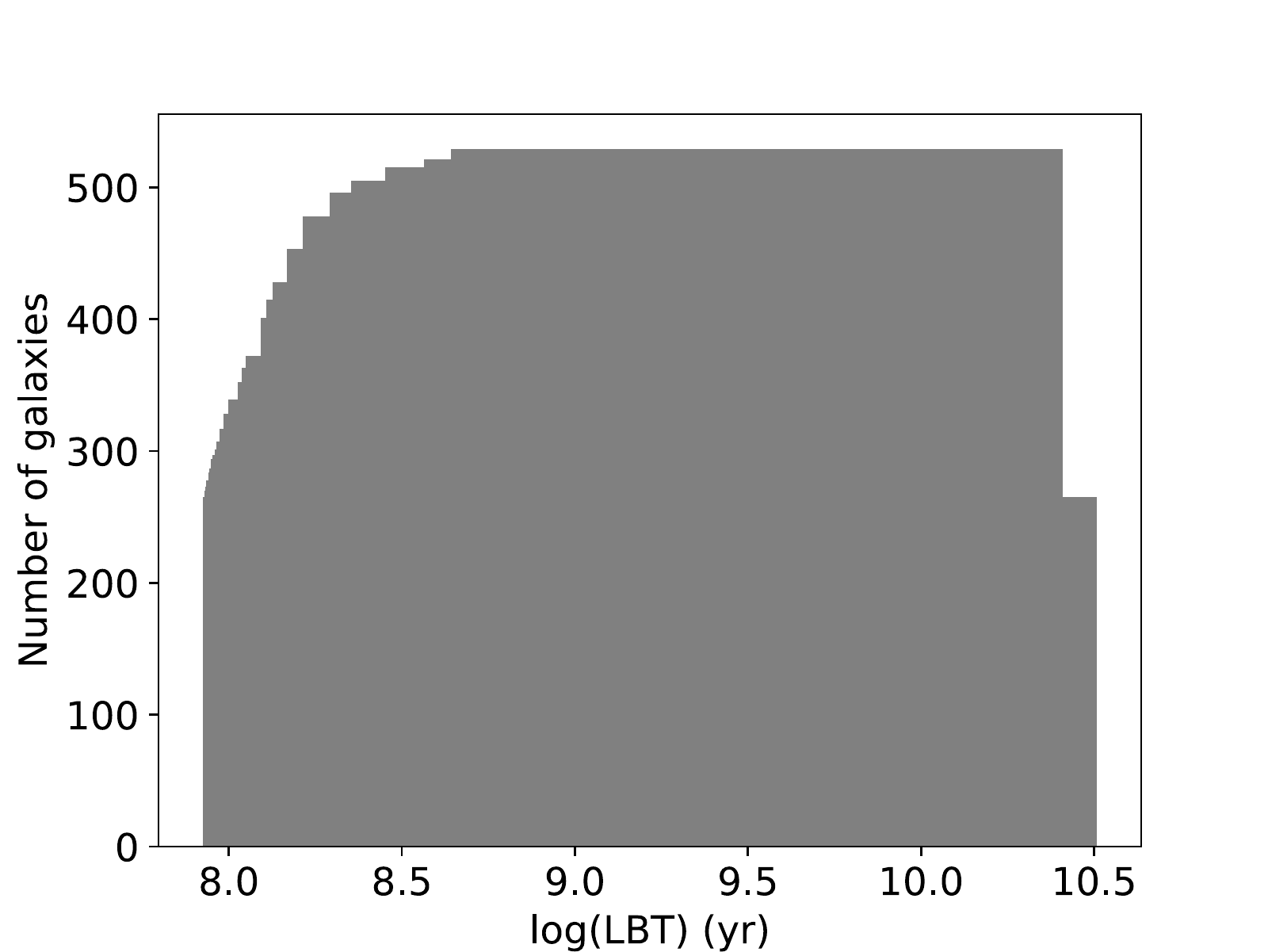}
\caption{Histogram of the coverage of the median time-line, showing the number of galaxies that populate each time.}
\label{fig:coverage}
\end{figure}

In Fig. \ref{fig:coverage} we show the number of galaxies that populate each time. Above $\sim$400 Myr ($10^{8.6}$ yr)  all galaxies are represented except in the very last value for the time. The minimum number of galaxies is 265, which is the number at both edges.

After interpolating all the ChEHs into the same timeline we can proceed to average them. We group the galaxies into mass, morphology and star-forming status bins (or any combination of these) in order to see how these parameters affect the ChEHs.

The averaging is conducted in two steps. First, we calculate the mean value of the ChEH for each galaxy over the time range where all galaxies within a particular bin have measured values of the metallicity. We take the average of these values and then shift all ChEHs by the difference, which in general terms bundles the ChEHs together. After this we perform the average of the shifted ChEHs. The reason we perform this is to prevent outliers from affecting the sample too much. As an example, consider the case where one galaxy in a bin has a very low metallicity and its high redshift meant its individual ChEH was not sampled for much of the recent time-line. In this case if we performed the average normally the galaxy would noticeably affect the metallicity only for a portion of the timeline, artificially creating what would appear to be a growth in metallicity for the bin in question. By shifting them beforehand the effect of any outlier is applied to the averaged ChEH over the whole timeline instead, preventing the appearance of such artefacts.

Performing the averaging in this manner has an additional advantage: we can separate the variance of the ChEHs within a bin of galaxies into two types: The variance due to the separation between individual ChEHs and the variance due to the different shapes of the ChEHs. This constitutes an additional result which allows us to infer whether galaxies in a bin have similar evolutions and whether they have a wide range of terminal metallicity values.

We also want to obtain the evolution of the MZR, which we derived from the ChEHs. But this time we used different mass bins in order to obtain a valid sample of the MZR. Then we can invert the relation between metallicity, LBT, and mass by making "vertical cuts" across the ChEHs. Thus, we select different times and obtain the metallicity for each mass bin. This immediately produces the mass-metallicity relation graphs at different times.

\begin{figure}
\centering
\includegraphics[width=\linewidth]{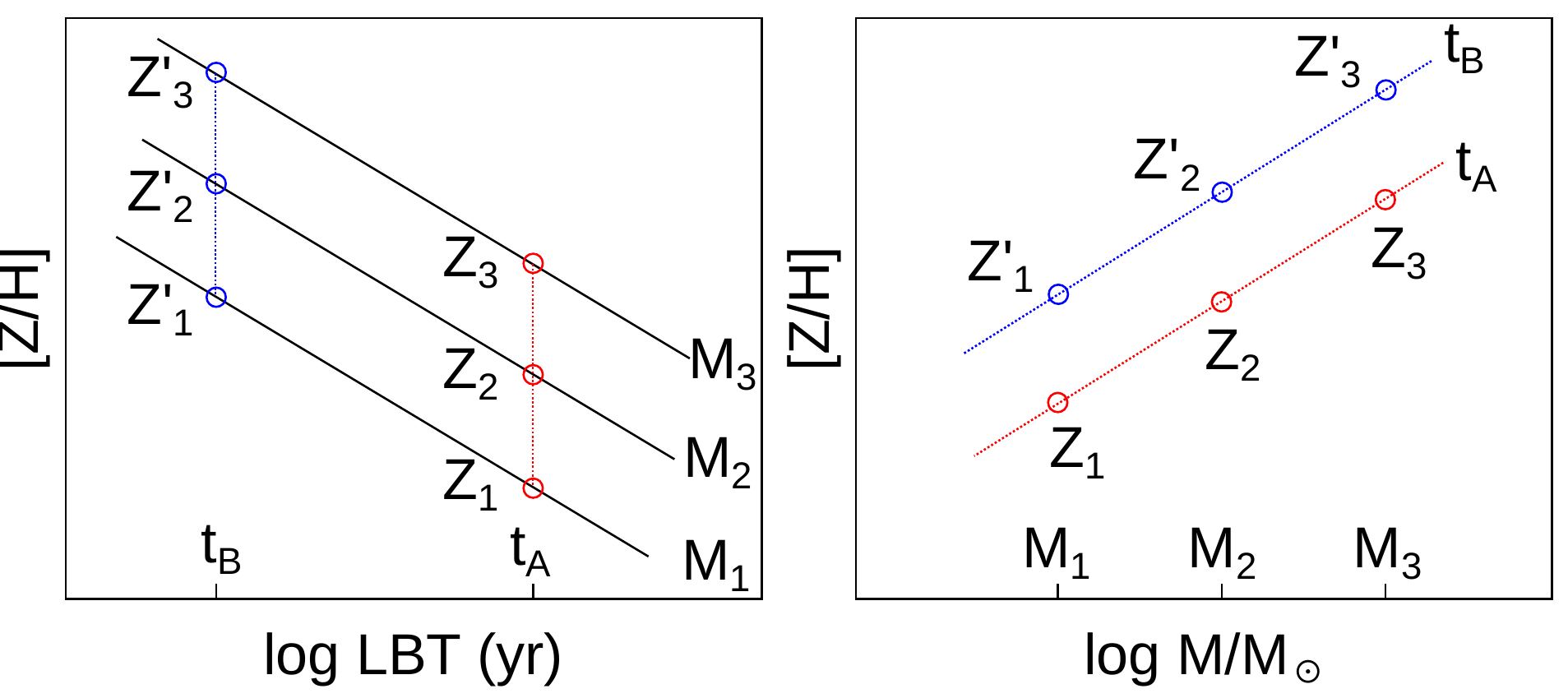}
\caption{Illustration of the process to obtain the MZR from the ChEHs. On the left panel we show a mock set of three ChEHs corresponding to three mass bins (M$_{1-3}$), on which we note the metallicities measured at two different times (t$_A$: Z$_{1-3}$, t$_B$ : Z'$_{1-3}$). On the right panel we show two MZR measured at two different times (t$_A$, t$_B$) showing the same mass-metallicity data points as in the left panel. This shows how we can obtain the MZR from the ChEHs by sampling a set of ChEHs vertically and converting the values into the mass-metallicity parameter space.}
\label{fig:cutting}
\end{figure}
In Fig. \ref{fig:cutting} we show an illustration of the process, where three ChEHs for three mass bins are used to obtain the MZR at two different times.

The mass bins used for this calculation are defined as being 0.5 dex wide with a 0.25 dex step. This means that half of each bin overlaps with the previous bin and half with the next one. This gives a smoother transition between mass values.
As a result of how the MZR is calculated, the mass values are those of that are currently measured for each galaxy, i.e., they correspond to the LBT of their redshift.

The last parameter studied in this article is the mass variation, which we use as a proxy for the star formation rate (SFR). It is obtained from the changes in mass divided by the time step (essentially dM$_*$/dt). The mass variation has two contributions, the increase in mass due to star formation and the decrease due to dying stars. The latter is a minor, steady decrease that does not contribute much to the mass variation ($\sim5\%$) which justifies our use of the mass variation as a proxy for the SFR. We indicate this quantity as SFR' in the article.

The uneven sampling of the LBT means that at recent times the time-steps are smaller and therefore the SFR' is more affected by the errors in the determination of the M$_*$ and SSP age, a result of dividing small increments in mass over small increments in time. In order to minimise this effect we fit a non-smoothing spline to the cumulative mass function and then take its numeric derivative as the mass variation. This mitigates the effect of the errors associated with the resolution.

\begin{figure*}
\centering
\includegraphics[width=\linewidth]{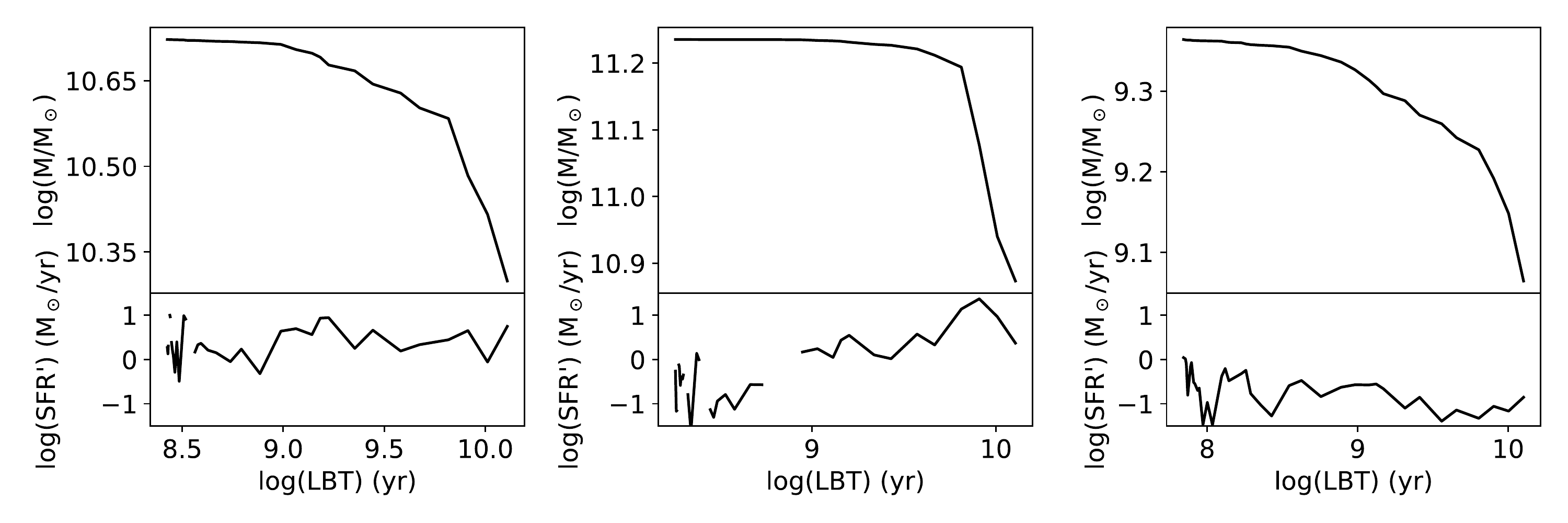}
\caption{Three examples of the mass evolution (above) and the resulting mass variation (below). At recent times the divergence of the mass variation is clearly observed. The gaps in the lines correspond to values of 0, which diverge upon taking the logarithm.}
\label{fig:dm_example}
\end{figure*}
In Fig. \ref{fig:dm_example} we show the mass evolution and SFR' for three galaxies. It can be seen that at recent times the SFR' oscillates, a result of the aforementioned small increments in time and mass. In order to average SFR' within a particular bin we calculate the individual SFR' curves in the same manner as we did for the ChEHs and then proceed similarly, first re-sampling them into a single timeline and then averaging them within the defined bins. The evolution of the SFR' shown here is compatible with the results for the evolution of the SFR depending on mass shown in previous studies \citep{Panter2007, Lopez-Fernandez2018, Sanchez2019}, which show how the more massive galaxies evolved faster with high SFR at early times that have declined, while the less massive galaxies show flatter profiles.


\section{Results}
In this section we present the results for the evolution of the metallicity and the MZR, obtained from the fitting of stellar population templates to each galaxy in our sample as described above. The metallicity we present here is a weighted average of the metallicities within an LBT bin over the galaxy and measured at the effective radius. 
All of the metallicity measurements shown here correspond to the stellar metallicity in the galaxies, not the nebular gas metallicity, which cannot be measured using our methodology to obtain the temporal evolution of the galaxies' properties.
One can weight the metallicity of the populations via either their luminosity or their mass. All the following results correspond to the metallicity weighted by luminosity. In general terms, weighing by mass or luminosity show similar trends in terms of evolution, but the trends are clearer for the luminosity weighed case as it is affected more by the younger populations.

Do note that the LBT we show here is not associated with a cosmological model. The "clock" that we are using is the isochrones that are fitted to stellar spectra to produce the SSP templates we use to analyse the data.

\subsection{Evolution of the metallicity}
\begin{figure}
\centering
\includegraphics[width=\linewidth]{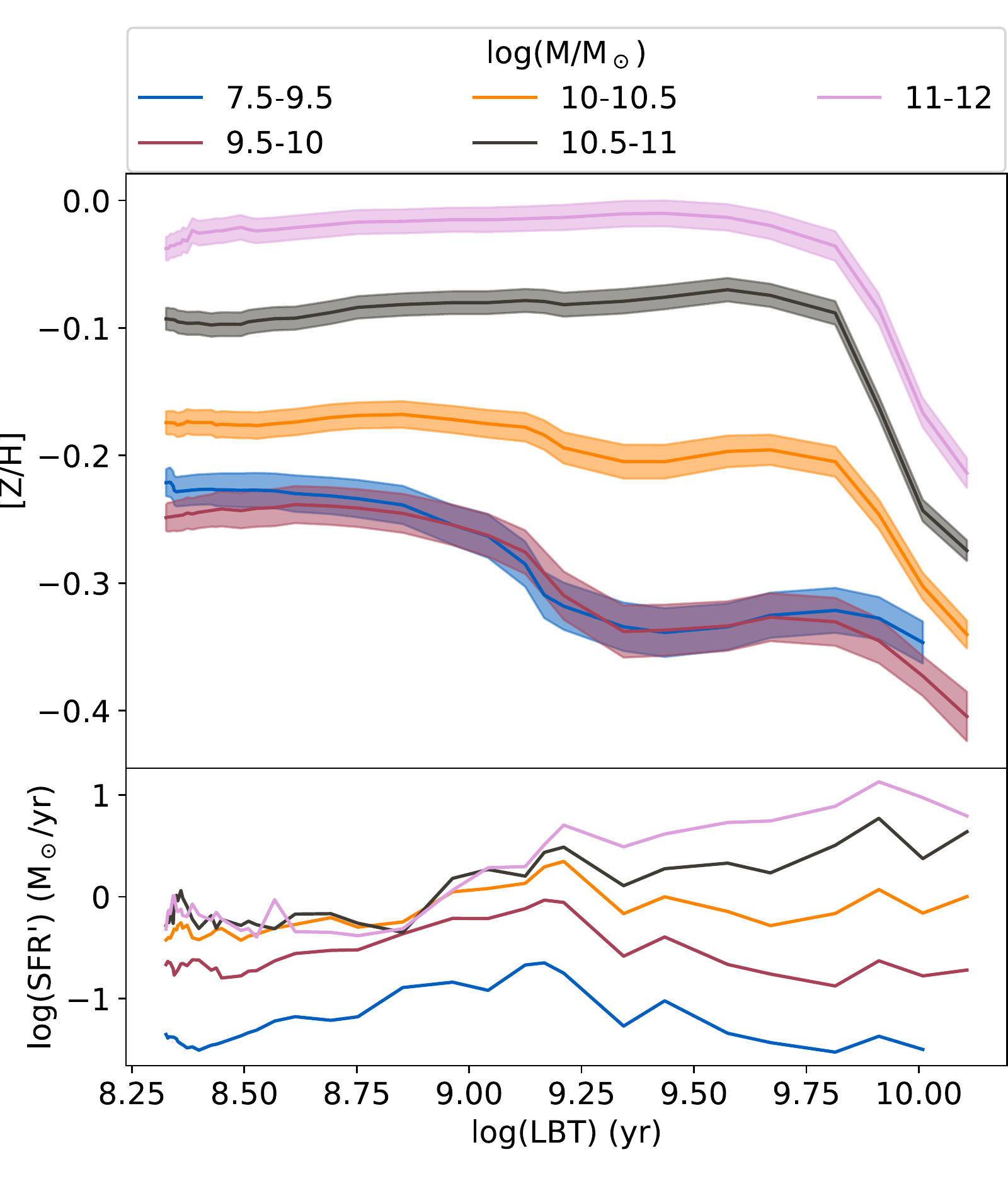}
\caption{On top, the evolution of the stellar metallicity [Z/H] as a function of the look-back time (LBT) for all the galaxies in the sample. Each line corresponds to a mass bin within which all the individual galaxies were averaged. The masses for the bins correspond to the mass as currently observed. The shaded areas represent the error of the mean ($\epsilon = \sigma / \sqrt{N}$), spanning the range between $Z/H - \epsilon_{Z/H}$ and $Z/H + \epsilon_{Z/H}$. The bottom panel shows the SFR' for the same mass bins.}
\label{fig:zh_all}
\end{figure}

In Fig. \ref{fig:zh_all} we show the average metallicity as well as the SFR' as a function of time for all the galaxies in our sample, with the error of the mean shown as shaded areas. The sample is divided into mass bins, depending on the currently observed value of the stellar mass for these galaxies. The most obvious feature is that the more massive galaxies are always more metallic than their lower mass counterparts. We can also observe that the more massive galaxies achieved their maximum metallicity (which in this case coincides with the current one) much earlier with the lowest masses still showing growth in metallicity. An apparent jump in metallicity is also observed between about $\sim$ 1-2.5 Gyr ($10^9$ - $10^{9.5}$ yr) in LBT. This jump is more prominent for the lower mass bins while not being visible for the highest mass ones.

The flatness of the profile for the more massive bins implies that for this population of galaxies the metallicity has reached its maximum value and they are no longer evolving from a chemical point of view. This lack of evolution can be partly explained as a result of the cumulative mass history of each galaxy, where each subsequent star-formation burst represents a lower percentage of the total mass of the galaxy being added. The relatively lower mass change then corresponds to a lower change in the global metallicity. Other than this, retired galaxies with no star formation will also not exhibit changes to their stellar metallicity regardless of the fact that their stars should still be enriching the ISM.

The fact that the lowest mass bins exhibit a steady growth up to current times implies that these galaxies are still assembling their mass and therefore are continuing to form stars within their increasingly enriched ISM. We could interpret this as low-mass galaxies having a slower evolution, whereupon the growth in metallicity observed here corresponds to a much shorter time-step for the more massive galaxies. In this scenario, if we were to continue observations for several Gyr more the growth in the metallicity of these galaxies would eventually taper off. Thus, producing a curve similar to that of the more massive galaxies but stretched in the temporal axis.

The dispersion is reflected by the error of the mean, an indication that while we can describe average trends the individual galaxies have much more complicated SFHs and ChEHs and therefore the averages are just a first order trend that has to be strongly modulated by the individual processes in galaxies. It is clear, however, that the error of the mean is smaller in more massive galaxies than for less massive ones. This implies that they have either more similar SFHs and ChEHs or that their differences appear at earlier times where we have less resolution. This is in agreement with the results by \cite{Ibarra-Medel2016}.

The SFR' curves show a clear negative evolution for the most massive bins, with a flatter, if bumpy, profile for the lower mass bins. The more massive galaxies also show higher values of the SFR' over time. The colours for the SFR' are equivalent to those of the ChEH, that is, a ChEH and a SFR' curve with the same colour within the same plot have been calculated using the same galaxies.

The most interesting feature of the SFR' curves, however, is the small peak in star formation that we find between 1-3 Gyr ($10^{9}$ - $10^{9.5}$ yr) in the LBT for all mass bins. This appears to occur immediately before the jump in metallicity that we detected in the ChEHs, and indeed they behave much in the same way, becoming more prominent for less massive galaxies.

\subsubsection{Effect of morphology}

In order to further evaluate how the properties of a galaxy affect its chemical enrichment we can separate the sample into morphology bins as described in section \ref{sec:sample}. Besides the final M$_*$, which has a clear influence on the SFHs in galaxies \citep{Perez-Gonzalez2008, Thomas2010, Perez2013, Gonzalez-Delgado2014b, Ibarra-Medel2016}, it is known that morphology strongly affects (or it is affected by) how galaxies evolve (\citealt{Blanton2009, Garcia-Benito2017, Sanchez2020} and references therein). In general we see that differences in morphology modulate the differences in evolution already observed as a result of galaxy mass.
It bears mentioning, however, that morphology is itself the result of evolutionary processes in the galaxies. When we say that the morphology affects the results what we are referring to specifically is that the evolutionary processes and properties which caused some galaxies to have a specific morphology also had a specific effect on the chemical evolution of those galaxies.

\begin{figure*}
\centering
\includegraphics[width=\linewidth]{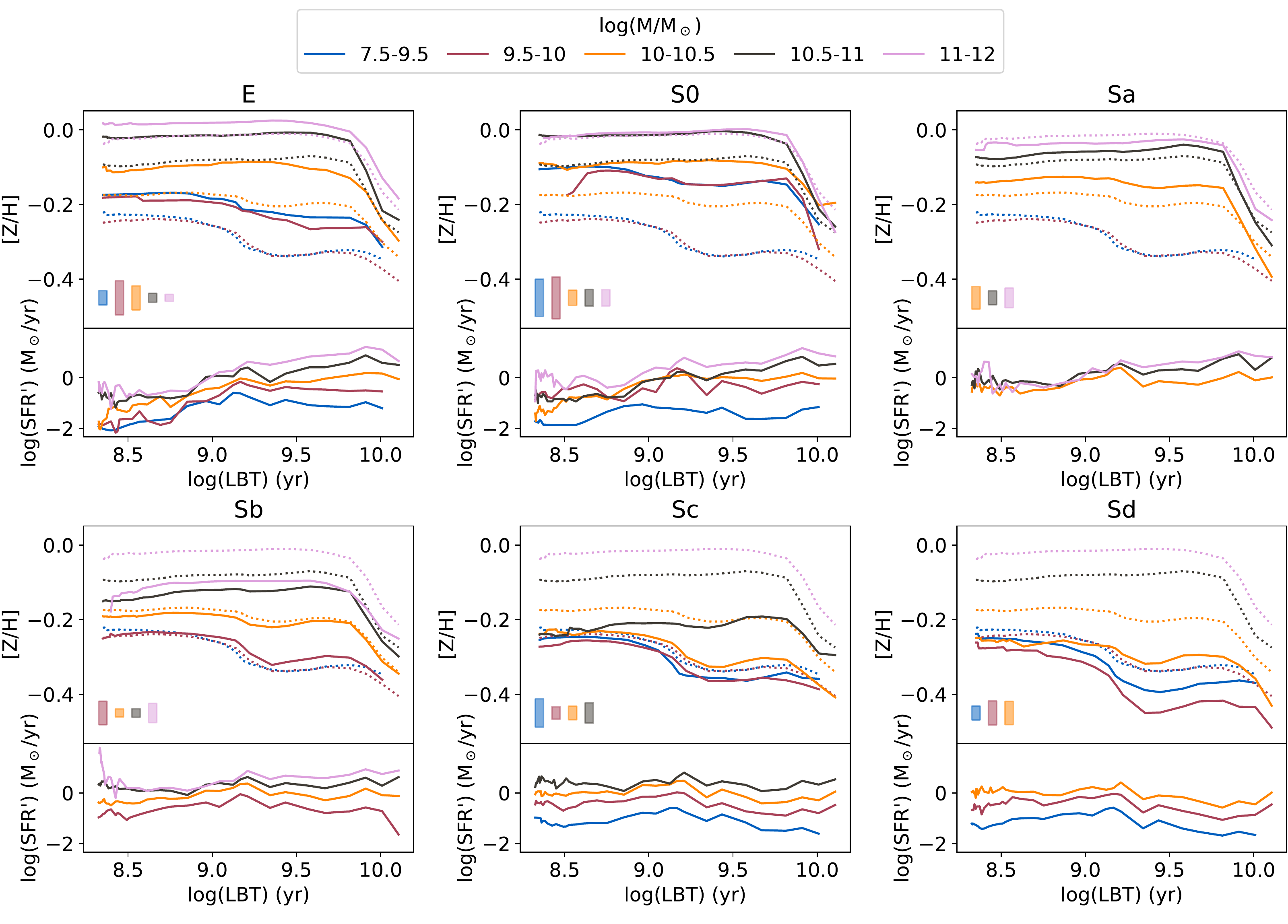}
\caption{In the top box of each panel we show the evolution of the stellar metallicity [Z/H] as a function of the LBT for each morphological bin of the galaxies in our sample. Each line corresponds to a mass bin for galaxies of the corresponding morphology, with the values for the mass-bins correspond to the mass as currently observed. The galaxies within each bin were averaged to produce each individual ChEH. The dotted lines represent the ChEHs for all galaxies (as in Fig. \ref{fig:zh_all}). In the lower left corner of the top box in each panel the shaded areas represent the typical error of the mean for each ChEH.
The bottom box of each panel shows the SFR' for the corresponding bins, in the same colours.
}
\label{fig:zh_morph_all}
\end{figure*}

In Fig. \ref{fig:zh_morph_all} we show the evolution of the stellar metallicity as well as the SFR' for the galaxies in our sample separated by morphology bins for different mass ranges. Within each box we also show in dotted lines the ChEHs for the entire sample (also shown at the top-left panel) for easy comparison. This allows us to quickly assess how the chemical evolution of a particular type of galaxies changes with respect to the full sample.
Showing the errors of each bin as we did in Fig. \ref{fig:zh_all} would not allow us to compare the ChEHs for each morphological bin with the full sample as the figures would be too loaded, so we indicate the typical error of the mean in bars in the lower left corner.

Early-type galaxies (E/S0) have consistently the highest metallicity values.
This is especially true for the more massive bins which are more representative of the sample for these morphologies and have a higher number of objects, as can be seen from their lower error bars.
This effect is also readily visible for spirals, with a clear trend from Sa to Sd galaxies of decreasing the metallicity, as well as a narrowing of the differences between different masses. Massive spirals also appear to exhibit a slight decrease in metallicity after the initial enrichment. However, it is fairly small compared to the uncertainties and therefore might not be statistically significant.

The earliest galaxy types also have substantially flatter profiles after the initial enrichment when compared to the later ones, which have increasingly steeper curves, especially for the Sc and Sd morphological bins. If we compare the low-mass bins for these morphological types with those for Sa and Sb (even for those in the E/S0 bin) we can see that the ChEHs get progressively less flat, switching from a flat evolution to a positive increase. This effect can also be seen for the entire sample, but Sd galaxies show a much more pronounced increase in the lower mass bins compared to the entire sample or any other morphological type.

The ChEHs for spiral galaxies exhibit an interesting feature: for those late type galaxies that populate all mass bins, the gaps between the curves for different masses get progressively narrower from Sa to Sc.
This appears to imply that for spirals the later the morphological type the less is the difference in metallicity between galaxies as a result of their differing masses. This effect should also be apparent in the MZR(t) as a progressive flattening towards later types.

Another feature that appears to be affected by the separation into morphology types is the aforementioned jump in the ChEH. As was mentioned for Fig. \ref{fig:zh_all} the jump is higher for lower mass bins. However, here we also see that it gets more prominent within a mass bin for the latest morphological types. Whereas in Sa it can only be slightly detected for the lower mass bins it gets deeper as we progress to Sb, Sc, and finally Sd, where it is very prominent.

This feature correlates with the SFR' of the galaxies as was observed in Fig. \ref{fig:zh_all}. The latter morphology bins show clearer peaks in SFR' compared to their earlier counterparts. The correlation is less clear in morphology than it is in mass, for example Sa galaxies show clear peaks in SFR' even though the jump in [Z/H] is barely visible only for the lowest mass bin. However, the general trend of both features being more prominent for later and less massive galaxies can still be observed. The fact that the earliest type galaxies such as E and S0 do not exhibit the bump in metallicity, as well as the less prominent bump in SFR', can be explained as a result of these galaxies being more likely to be retired, having formed most of their mass already. This would dilute the metallicity signature compared to the later morphological bins.

The results appear to suggest that at around redshift 0.1-0.2 (1-2.5 Gyr ago) there was a general episode of star formation which affected the metallicity of galaxies in the local universe. It is not required for all galaxies to have a peak in star-formation in order to produce this feature, it is enough if it appears on a significant portion or if there is a tendency to have somewhat higher SFR than normal around that time. It does, however, need to be a phenomenon of a global nature rather than some outliers affecting the average, as the feature appears in all the mass bins we consider for our sample with a similar appearance and at the same time.

This episode of star-formation has not been reported in observed galaxy surveys (to our knowledge). However, this means little, since our procedure is recovering the SFH and ChEHs of individual galaxies, not of a population. For instance, a morphological or mass change from the current time to the event time may dilute this signal in a high-z survey. On the other hand, if it was an artefact of the procedure, it is difficult to justify why it shows such a clear morphology/mass pattern.

\subsubsection{Effect of star-forming status}
\label{sec:zht_sfs}
Another parameter that can affect the chemical evolution of galaxies is whether or not they are still forming stars. Star-forming galaxies (SFG) have either sustained star formation (SF) throughout their life-times or they have experienced a massive recent burst of SF that rejuvenated them. Recent results, however, give more support to the first scenario \citep{Pandya2017, Sanchez2019} showing that galaxies currently in the star-forming main sequence have mostly stayed in that regime or oscillated in and out repeatedly through time.

This means that we would expect these galaxies to still be evolving in terms of their metallicity.
"Retired" galaxies (RG) have stopped forming new stars at the current observing time and therefore have an aged stellar population with no further evolution in their stellar metallicity. The current population of stars in retired galaxies continues to enrich the ISM, but without a new stellar population formed from this enriched gas, the stellar metallicity remains unchanged. Furthermore, it is not clear when these galaxies stopped their SF activity (and the subsequent stellar metal enrichment). Some recent studies support the idea that a substantial portion of those galaxies were SF $\sim$1-4 Gyr ago \citep{Muzzin2013, Rodriguez-Puebla2017, Sanchez2019}.

\begin{figure*}
\centering
\includegraphics[width=\linewidth]{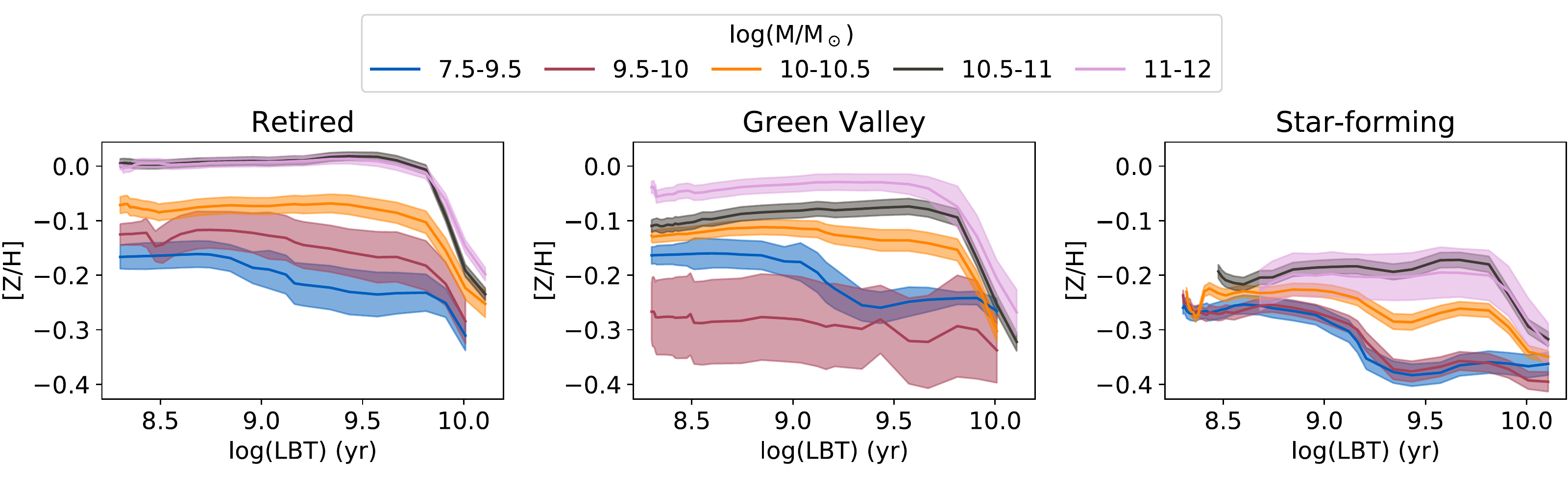}
\caption{Evolution of the stellar metallicity [Z/H] as a function of the look-back time (LBT) for all the galaxies in our sample, separated by their star-formation status. In the left-panel we show the retired galaxies, in the middle one those in the Green Valley, and finally in the right-panel the star-forming galaxies. Each line corresponds to a mass bin within which all the individual galaxies within the corresponding star-forming status were averaged. The values for the mass-bins correspond to the mass as currently observed. The shaded areas represent the error of the mean for each ChEH.
}
\label{fig:zh_sf}
\end{figure*}

In Fig. \ref{fig:zh_sf} we show the evolution of the stellar metallicity for galaxies in our sample separated by their current star-forming status. We defined the SFG bin as galaxies with an average EW$_{\mathrm{H\alpha}}$ over 10 \AA{} while RGs were defined as those with EW$_{\mathrm{H\alpha}}$ below 3 \AA, leaving the Green Valley galaxies (GVG) as those between the two aforementioned values.
These values were chosen to be consistent with the sample selections of \cite{Lacerda2020}. \cite{Cano-Diaz2019} show that 5-6 \AA{} is a good value to separate a sample into SFG-RG bins as shown from the object density plots in the SFR-M$_*$ diagram. The values we use allow us to define the GVG bin of galaxies while also ensuring that the objects contained in the SFG and RG bins are robust identifications.

\begin{table}
\centering
\caption{Distribution of galaxies in the sample regarding their star-forming status}
\setlength{\tabcolsep}{5pt}
\begin{tabular}{lccc}
\hline
	&	RG	& GVG & SFG	\\
Total & 218 & 81 & 230 \\
E	&	122 & 7 & 2 \\
S0	&	63 & 4 & 6\\
Sa	&	21 & 23 & 9 \\
Sb	&	9 & 42 & 88 \\
Sc	&   1 & 3 & 69 \\
Sd  &	2 & 2 & 56 \\

\end{tabular}

\textbf{Notes.} Table showing the number of galaxies in each star-forming status in total and divided by morphology. RG = Retired Galaxies, GVG = Green Valley Galaxies, SFG = Star-Forming Galaxies.
\label{tab:sfs}
\end{table}	

In Table \ref{tab:sfs} we show the number of galaxies in each star-forming status bin, with further detail as to the morphological makeup of each bin.

It is important to mention that the SFG-GVG-RG separation is not independent of the separation by morphology. SFG are predominantly late-type galaxies while RG are mostly early type E and S0 galaxies (e.g. \citealt{Young1996, Sanchez2007, Cano-Diaz2019}) which can be clearly be observed in Table \ref{tab:sfs}. As such, the main results observed from separating the galaxies by morphology still apply: The RG have higher metallicity values and were enriched faster than their SFG counterparts, with the GVG between them. The GVG, however, appear to be closer to the RG in their chemical evolution than to the SFG.

An interesting new feature can be observed in the SFG bin, in that the current metallicities for all mass bins show an apparent tendency to converge. This is opposite to what we observe for the GVG and the RG which maintain a gap between the metallicity of their mass bins. The SFG do appear to have a wide gap between bins initially, but it then steadily narrows.

This suggests that some form of regulatory process occurs in the SFG which limits the maximum value the metallicity can reach, thus as galaxies evolve they converge around that value. The two most massive bins have mostly flat curves with a metallicity value substantially lower than that of the other bins, while the lower mass bins have increased their metallicity steadily but are now flattening around the same value.

As mentioned above, low mass galaxies tend to be still becoming enriched in metallicity, which is observed here mostly for the SFG with a steep growth, but it is also observed in the GVG and the RG albeit in a much more subdued manner. This phenomenon is similar to what is observed comparing the early to late type galaxies, so it is likely to be a result of morphology.

The RG show a clear progression in M$_*$ with all the ChEHs having a similar shape but with an initial enrichment that ends earlier with increasing mass. This is a clear result of downsizing, with the more massive galaxies assembling their mass faster.

\subsection{Evolution of the mass-metallicity relation}
\label{sec:mzr_evo}
In the previous section we analysed how the values of the stellar metallicity have evolved in cosmic time for the galaxies in our sample. In this section we will perform a similar analysis for the MZR.
We can use the parameters obtained from our analysis to explore how this relation has changed throughout time for the galaxies in our sample.

It is expected that galaxies with higher mass both achieve higher values of the metallicity and evolve faster, a fact that could already be observed to an extent in the previous section. 
On the contrary, galaxies in the lower mass bins appeared to still be in the process of increasing their metallicity.

A faster evolution of higher mass galaxies should manifest in the MZR as a change in slope over time, with the relation being steeper in earlier times as the more massive galaxies should have already been enriched while the less massive galaxies would be much slower. Then, as time passes, the relation gets shallower as the lower mass zone increases its metallicity while the high mass galaxies maintain the values from earlier times. Of course, the delay in enrichment for lower mass galaxies is not the only possible cause for a slope change. Even if all galaxies reached their maximum metallicity at the same time, a difference in the maximum value of the metallicity depending on mass would still produce a change in slope over time.

\begin{figure}
\centering
\includegraphics[width=\linewidth]{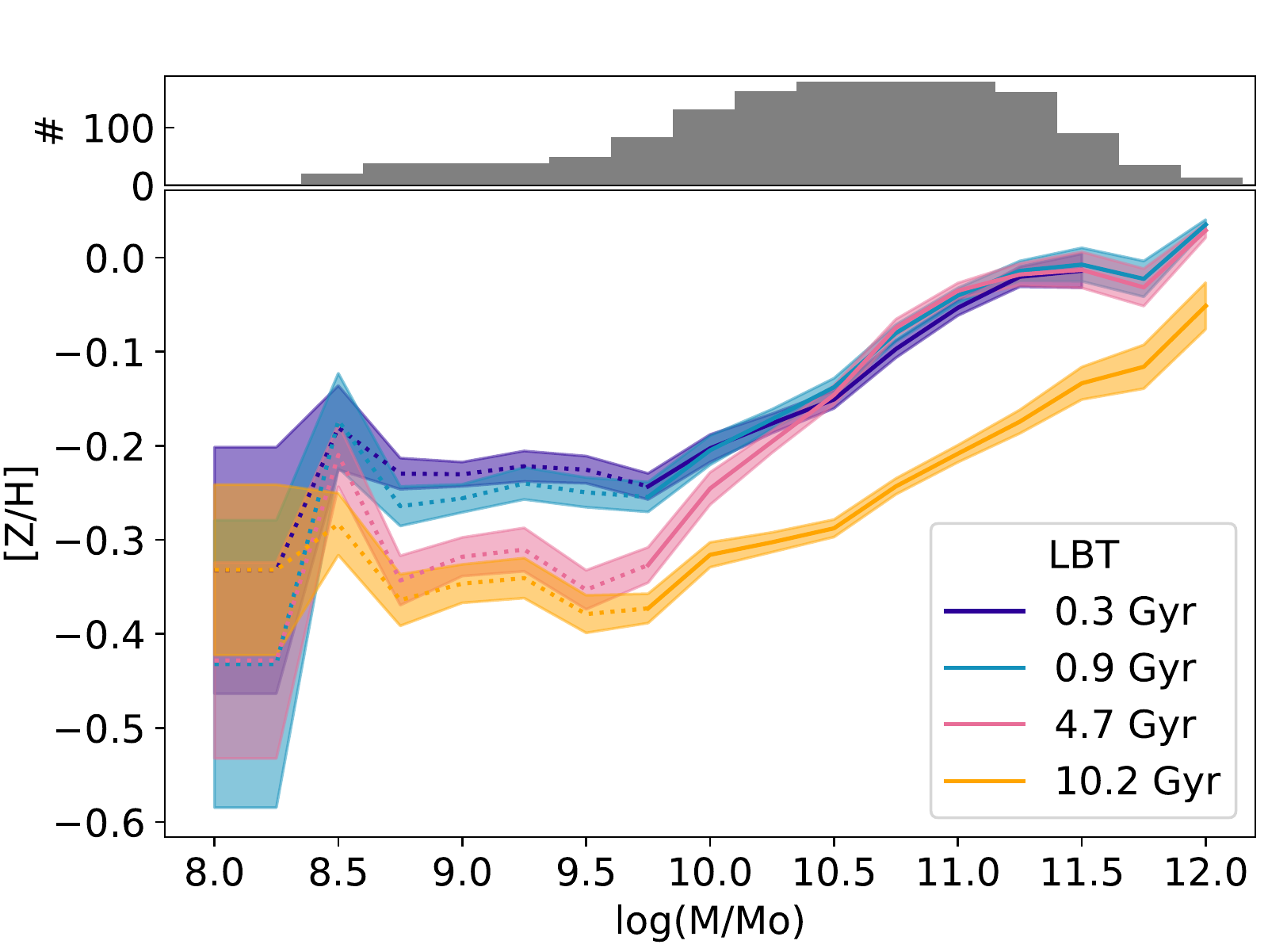}
\caption{Evolution of the stellar MZR in time for all the galaxies in the sample. Each line represents the MZR at a particular look-back time (LBT) as indicated by the colour. The shaded areas represent the error of the mean ($\epsilon = \sigma / \sqrt{N}$), spanning the range between $Z/H - \epsilon_{Z/H}$ and $Z/H + \epsilon_{Z/H}$. In the top panel we show the distribution of galaxies as a function of the mass range. Note that due to the way we have calculated the MZR the mass values always correspond to the currently observed mass (see Sec. \ref{sec:averages}).}
\label{fig:mzr_all}
\end{figure}

In Fig. \ref{fig:mzr_all} we show the MZR for several values of the LBT for the galaxies in our sample. The expected delay in enrichment is clearly observed: the higher mass galaxies quickly get enriched, which initially makes the slope steeper. On the contrary, the lower mass galaxies get enriched much slower, steadily making the slope shallower once more, but at higher overall values of the metallicity. Note that the stellar mass values do not change with time, they correspond to the currently observed properties of the galaxies. This can be inferred by how the mass coverage does not vary with the LBT.

The lowest LBT bin in the figure does not cover the highest values in the considered mass range. This is a result of the fact that the few most massive galaxies in our sample are not in our immediate vicinity, so as a result of having a higher redshifts their lowest LBT value is higher than the lowest LBT bin. In other words, there is no overlap between very high mass and very low redshift values for the galaxies in our sample. This is the result of an unavoidable selection effect: the higher mass galaxies (e.g., Andromeda) also have a larger size, so if they were placed close to us they would not fit within the field of view of the instrument. Thus, this is a direct effect of the diameter selection of the sample, aimed to fit its optical extension with the FoV of the instrument. However, other selections, such as the one proposed for the MaNGA survey with a wider range of redshifts \citep{Bundy2015, Wake2017}, do indeed produce an even stronger M$_*$-redshift effect, as highlighted by \cite{Ibarra-Medel2016}.

The MZR for those mass regions outside the range in which the CALIFA galaxies is considered to be complete \citep{Walcher2014} are represented with dotted lines.
For the regimes whose completeness we cannot guarantee the number of sampled galaxies is relatively low (Fig. \ref{fig:mzr_all}, top panel). The combination of the two effects prevents us from drawing strong conclusions about this regime. In particular, the results for the range of low-mass/dwarf galaxies should be taken with particular caution.

Consistently with the previous caveat, it is observed that the mass range below about $10^{8.7}$ solar masses has very high fluctuations in the value for the metallicity, with similarly high errors. This is a consequence of the small number of galaxies in our sample in this range with only a few objects.

\subsubsection{Effect of morphology}
\label{sec:mzr_evo_morph}
As we did for the evolution of the metallicity we can divide our sample into morphology bins and calculate how the MZR evolves within each of them. Owing to the aforementioned caveat of a very small number of galaxies below $10^{8.7}$ solar masses, which already made results for the full sample in this range unreliable, the MZR below this value was removed in the morphology bins. We also removed any mass-metallicity data point that did not have at least 3 galaxies contributing to it.

\begin{figure*}
\centering
\includegraphics[width=\linewidth]{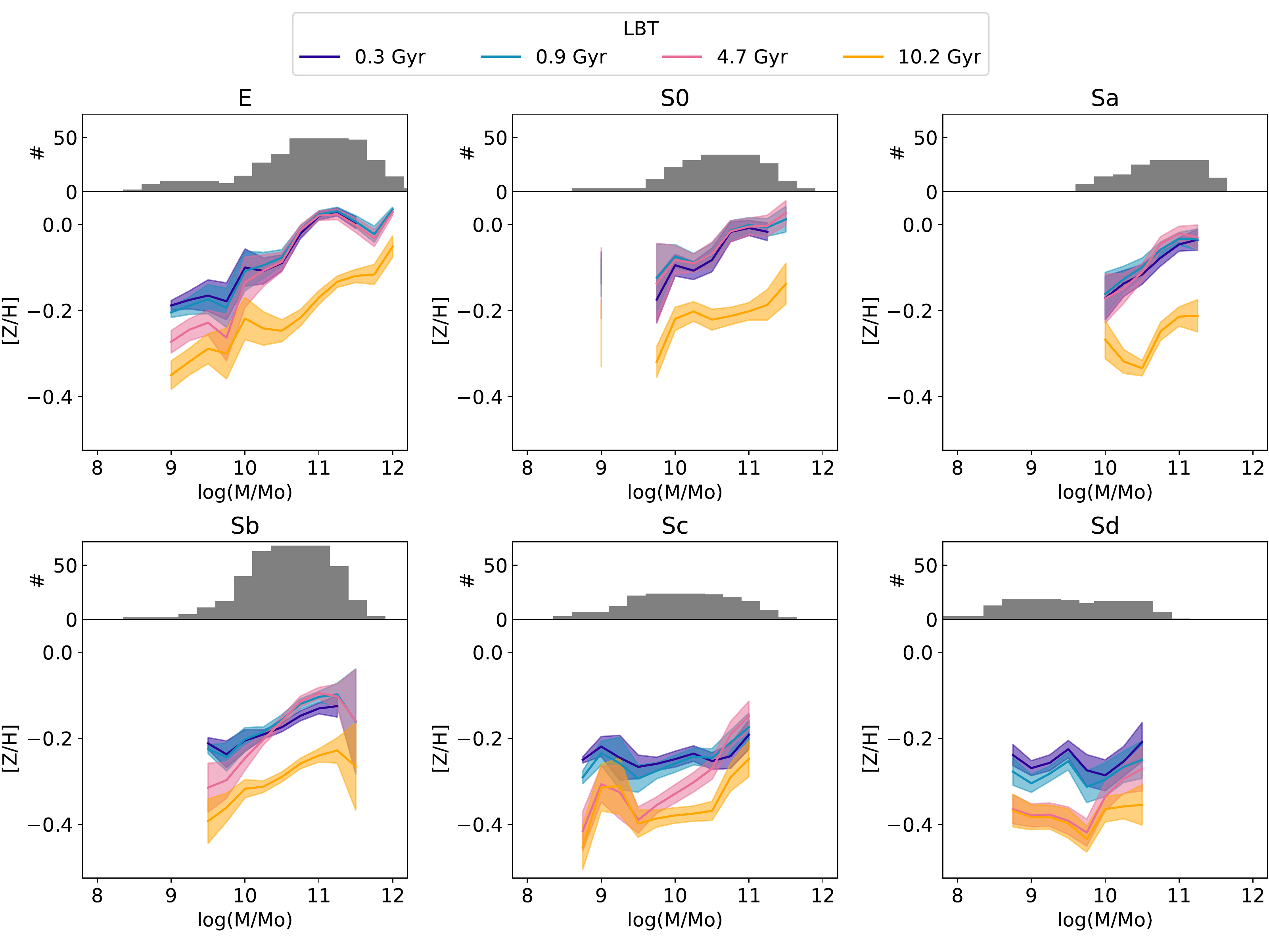}
\caption{The bottom box of each panel shows the evolution of the stellar MZR in time for the galaxies in our sample separated into morphological bins, with the shaded areas representing the error of the mean within that sample. On the top box of each panel a histogram showing the distribution of galaxies with the mass range is shown. Note that due to the way we have calculated the MZR the mass values always correspond to the currently observed mass (see Sec. \ref{sec:averages}).
}
\label{fig:mzr_morph_all}
\end{figure*}

In Fig. \ref{fig:mzr_morph_all} we show the MZR calculated for each morphological bin. The first thing that can be observed is that not all the boxes have the same coverage in mass. Indeed, this is a result of the fact that the earlier types of galaxy tend to be more massive than the later types, though E galaxies cover a very wide range in stellar mass. The extremes of this can be appreciated by comparing S0 to Sd, which have a very small overlap in their mass range.

A remarkable result that can be observed as we divide the sample into morphology bins is that half of them do not exhibit a delay in the chemical enrichment for lower masses. Sc is the only morphology bin to show a delay in enrichment like the one observed in Fig. \ref{fig:mzr_all}, with E and Sb galaxies showing some delay for their lowest mass values only. S0, Sa and Sd clearly show a steady enrichment where the mass only determines the values of the metallicity, not how fast it grows.

It appears then, that the delay in enrichment that is visible for the full sample cannot be attributed solely to the mass of the galaxies. Indeed, the shape of the MZR for the full sample coincides with what we would obtain by combining the earliest types (E and S0) with the latest (Sc-d) morphological types. The correlation between mass and morphology, with the later types dominating the low mass regime and the early types the high mass regime would be sufficient to explain the evolution of the MZR for the full sample.

This implies that morphology is also a main actor regarding how a galaxy enriches itself. Mass determines the degree of enrichment a galaxy can achieve, and within certain morphological types it affects the rate of enrichment, but the global change in how quickly galaxies get enriched is mostly due to how morphology affects the ChEH.

As we progress to later types the MZR gets progressively flatter, earlier types have a strong dependence on the mass for the metallicity but later types less so. Indeed, Sd galaxies have a fairly flat MZR now and in the past, suggesting that they get enriched the same way irrespective of their mass.

\subsubsection{Effect of star-formation stage}
As we did for the evolution of the metallicity, we can separate our sample into bins depending on their star-forming status. The bins are defined as we did for the evolution of the metallicity, consistent with those identified in the star formation main sequence for galaxies.

\begin{figure*}
\centering
\includegraphics[width=\linewidth]{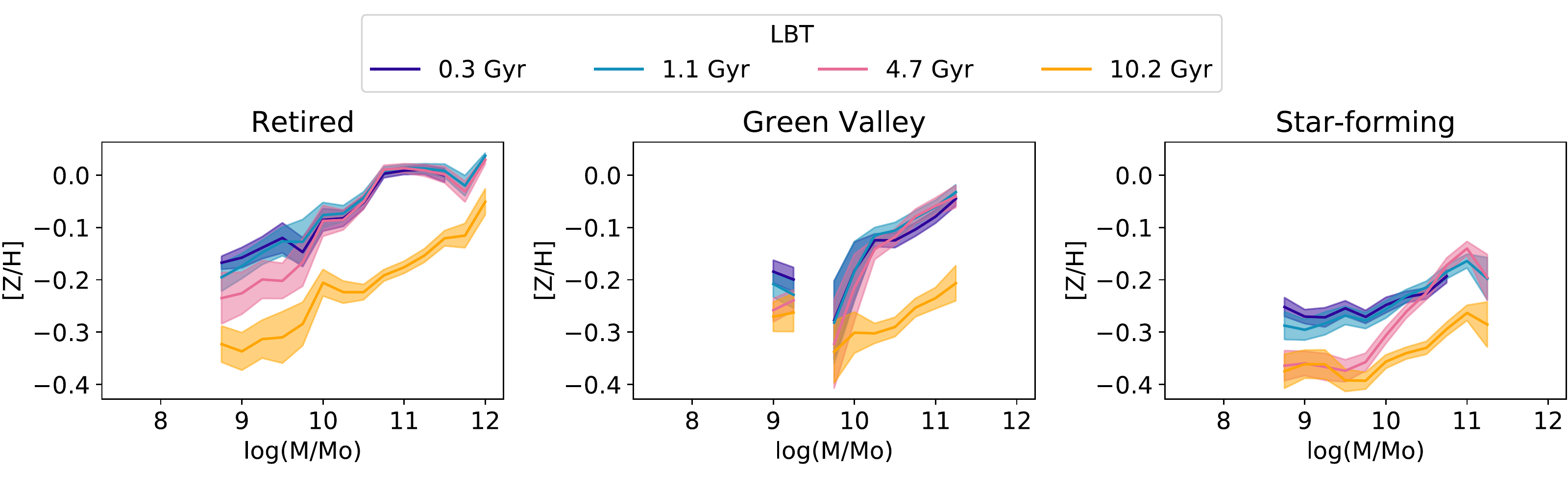}
\caption{Evolution of the stellar MZR in time for all the galaxies in our sample, separated by their star-forming status. On the left we show the retired galaxies, in the middle those in the Green Valley, and finally on the right the star-forming galaxies. Each line represents the MZR at a different LBT as indicated by its colour. The shaded areas represent the error of the mean ($\epsilon = \sigma / \sqrt{N}$), spanning the range between $Z/H - \epsilon_{Z/H}$ and $Z/H + \epsilon_{Z/H}$. Note that due to the way we have calculated the MZR the mass values always correspond to the currently observed mass (see Sec. \ref{sec:averages}).
}
\label{fig:mzr_sf}
\end{figure*}

In Fig. \ref{fig:mzr_sf} we show the MZR for the SFG, GVG, and RG galaxies. The SFG are generally less metallic than the GVG and the RG, and have a significantly flatter MZR, especially at recent times. The SFG and the RG show a delay in the enrichment for lower mass galaxies, though for the RG it only appears for the lowest masses in their range, similar to the elliptical morphology bin. The GVG, on the other hand, exhibit a clear change in slope but no delay in the enrichment, suggesting that mass strongly regulates the amount of metals that these galaxies produce and retain, but it has no effect on the rate of enrichment.

In the previous section we discussed how it is possible that the chemical enrichment delay in the global MZR appears as a result of mixing galaxies of different morphology. The separation between the star-formation stages is not independent from the morphology: The RG area of the SFMS is mostly populated by early type galaxies, which in Fig. \ref{fig:mzr_morph_all} all have similar shapes of the MZR. It is to be expected, then, that they follow the behaviour of the E galaxies. The GVG are more of a mixed bag, but still much more likely to be earlier spirals than later types. The change in slope for the full population appears as a consequence of adding Sc and especially Sd galaxies. They clearly have a delayed chemical evolution for lower masses and much flatter profiles than the earlier types. These types dominate the lower mass portion of the sample, much as the earlier types dominate the higher mass portion. It is only in the SFG bin, then, that Sc-d galaxies make up a significant portion of the sample, which combined with earlier type, more massive galaxies, allows it to show the clear delay in enrichment of the global MZR.
See Table \ref{tab:sfs} in Section \ref{sec:zht_sfs} for the distribution of our sample in the RG-GVG-SFG bins as well as the morphologies within them.

\subsection{Dependence on radial distance}

So far we have been analysing global properties of the galaxies over time. A detailed study of the extended properties of galaxies throughout their lifetime can be ambiguous as we would need to assume that the population of stars we analyse at each spatial position has spent the entirety of their lifetime at that same position. It has been known for some time that stellar mixing and migration affects the metallicity distribution in discs, blurring the gradients and widening the scatter in the relations \citep[e.g.][]{Haywood2008,Schonrich2009, Minchev2010, DiMatteo2013, Sanchez-Menguiano2016, Martinez-Medina2016, Martinez-Medina2017, Sanchez-Menguiano2018}. 

Most of the current simulations of radial migration describe a general increase in the dispersion of the radial distributions rather than a global change in their profiles \citep{Minchev2010, DiMatteo2013}. In general, radial trends are slightly smoother and their signature mitigated or even blurred (in the extreme cases). Thus, any radial difference observed in the average properties using the fossil record could be considered as a lower limit to the one that would be observed without migration/mixing.

Additionally, the type of analysis we are performing is blind to the effects of mergers. If a galaxy in our sample is the result of a major merger in the past what we would observe is a sort of average of the properties of the two progenitor galaxies, weighted by their relative mass. This should have little effect on the results we are reporting in regard to the global properties, as those are already the result of an combining several galaxies.
For spatially resolved properties, however, mergers can affect the results as they are expected to perturb the orbits of stars significantly, driving them to different galactocentric radii than those at which they were born.

Even with the caveats described above it is interesting to probe how the properties measured at the centre of galaxies differ from those at the outskirts. Whether this is an effect of birth (describing the original distribution of where stars were born) or modulated by migration or mergers is not possible to disentangle with the current observations, but can be traced by detailed simulations (e.g., \citealt{Ibarra-Medel2019}).

\begin{figure*}
\centering
\includegraphics[width=\linewidth]{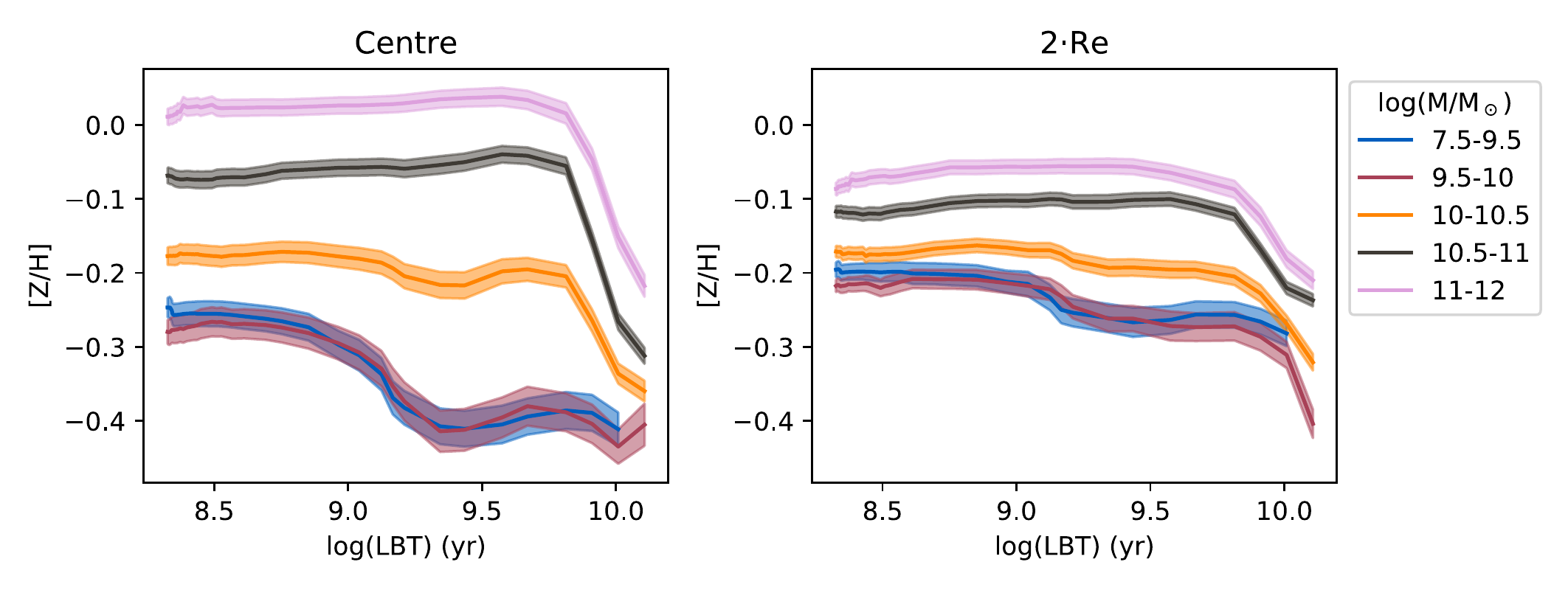}
\caption{Evolution of the stellar metallicity [Z/H] as a function of the look-back time (LBT) for all the galaxies in our sample, separated between whether the metallicity is measured at the centre or at two effective radii. Each line corresponds to a mass bin within which all the individual galaxies within the corresponding morphological bin were averaged. The mass bins correspond to the currently observed values of the mass for the galaxies in our sample.
}
\label{fig:zh_radius}
\end{figure*}

\begin{figure*}
\centering
\includegraphics[width=\linewidth]{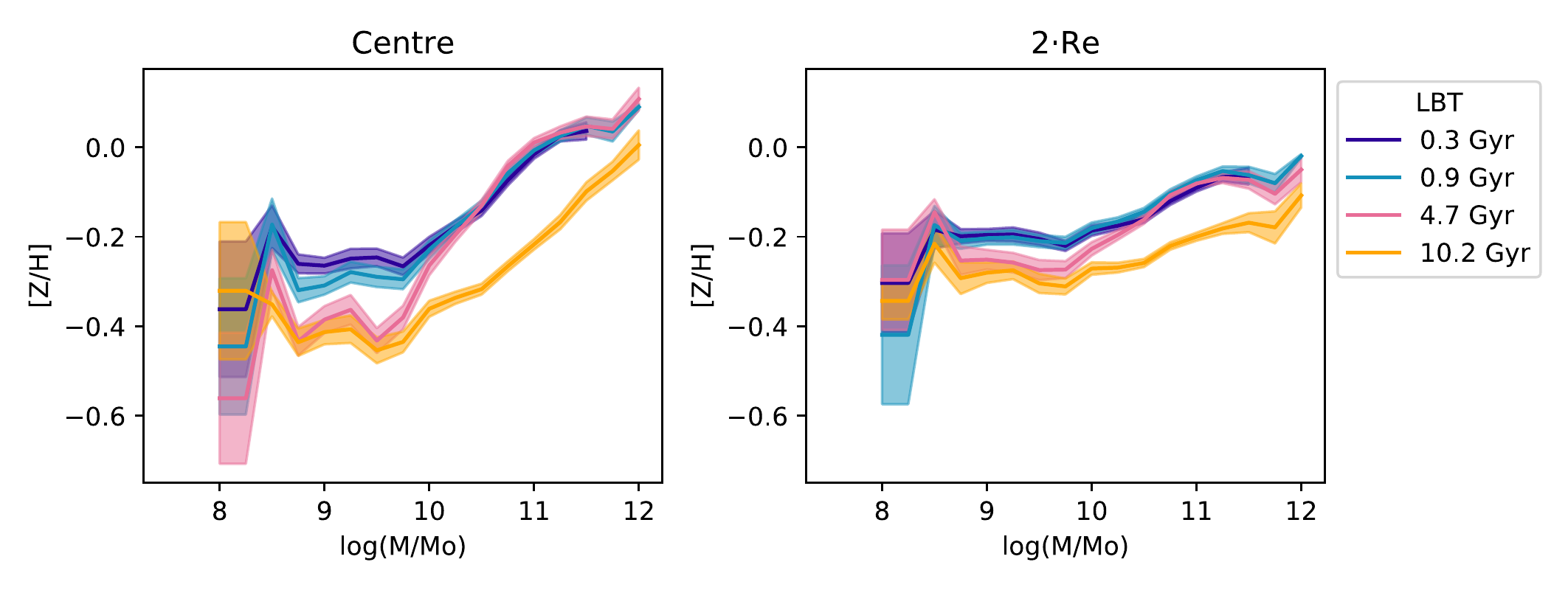}
\caption{Evolution of the stellar MZR in time for all the galaxies in our sample, separated between whether the metallicity is measured at the centre or at two effective radii. Each line represents the MZR at a different LBT as indicated by its colour. Note that due to the way we have calculated the MZR the mass values always correspond to the currently observed mass (see Sec. \ref{sec:averages}).
}
\label{fig:mzr_radius}
\end{figure*}

In Fig. \ref{fig:zh_radius} we show the ChEHs and in Fig. \ref{fig:mzr_radius} we show the evolution of the MZR, both measured at the centre and at two effective radii. The measurements for these quantities at the effective radius is shown in Figs. \ref{fig:zh_all}, \ref{fig:mzr_all}.

In both figures we can observe that the differences between the curves are more prominent at the centre compared to those at the effective radius. Conversely, the differences are much smaller at the outskirts. This applies both to the separation in mass and time.

The ChEHs become flatter and the range between the different masses becomes narrower around the ChEH for the $10^{10}$ - $10^{10.5}$ solar masses, which barely changes. This means that galaxies of higher masses become less metallic at the outskirts whereas galaxies less massive than this bin become more metallic compared to their centres. This shows how the less massive galaxies have a positive metallicity gradient (outside-in) and the more massive galaxies have a negative gradient (inside-out). The switch of the growth mode for low mass galaxies to outside-in has been reported in other studies (\citealt{Perez2013, Gonzalez-Delgado2017, Garcia-Benito2017, Goddard2017, Sanchez2020}).

The MZR also gets flatter at the outskirts, minimising the differences between galaxies of different mass. The change in time also gets less distinct, with little evolution and no differential evolution as a function of mass after the first growth in metallicity after the oldest MZR.

The fact that all galaxies are fairly similar at the outskirts can be understood to be the result of local downsizing \citep{Perez2013, Ibarra-Medel2016, Garcia-Benito2017}. Galaxies have generally exponential profiles in surface mass density, $\Sigma_*$ \citep{Gonzalez-Delgado2014a, Sanchez2020}. Thus there may be a large difference in $\Sigma_*$ in the centre but a much less pronounced difference in the outskirts. Local downsizing means that areas with similar $\Sigma_*$ have evolved at a similar rate, explaining the behaviour observed here. However, it could also be a consequence of the difference between the radial profiles of low mass galaxies (rising from the inside-outwards) and those of more massive ones (declining from the inside-outwards). It is readily apparent in Fig. \ref{fig:zh_radius} that the more massive bins have higher metallicities at the centre than in the outskirts, whereas for low mass galaxies the metallicity at the outskirts is higher than at the centre.

It should be mentioned that the gradient adopted to derive the inner- and outer-most metallicities is assumed to be linear (following \citealt{Sanchez2018}), so any deviations from this behaviour will introduce uncertainties in the values. This, however, is not expected to be a significant effect when averaging over several galaxies, and certainly should not affect the qualitative behaviour we are reporting here.

\subsection{The variance of the shape and gap between galaxies}

As described in Section \ref{sec:averages} the method we employ to perform our averages within a bin of galaxies allows us to separate the variance between ChEHs into two different sources: The variance due to having a different shape and that due to a gap between curves. Two galaxies with a large shape variance but a small gap variance will have a similar value of the metallicity averaged over the LBT, but very different chemical evolution histories. In the opposite case, two galaxies with a large gap variance but a low shape variance will exhibit practically the same chemical evolution histories but with different average metallicity values over that history.

This information allows us to probe whether galaxies in a particular bin are likely to have similar chemical histories and whether they have a large variance in terms of their current metallicity.

\begin{figure*}
\centering
\includegraphics[width=\linewidth]{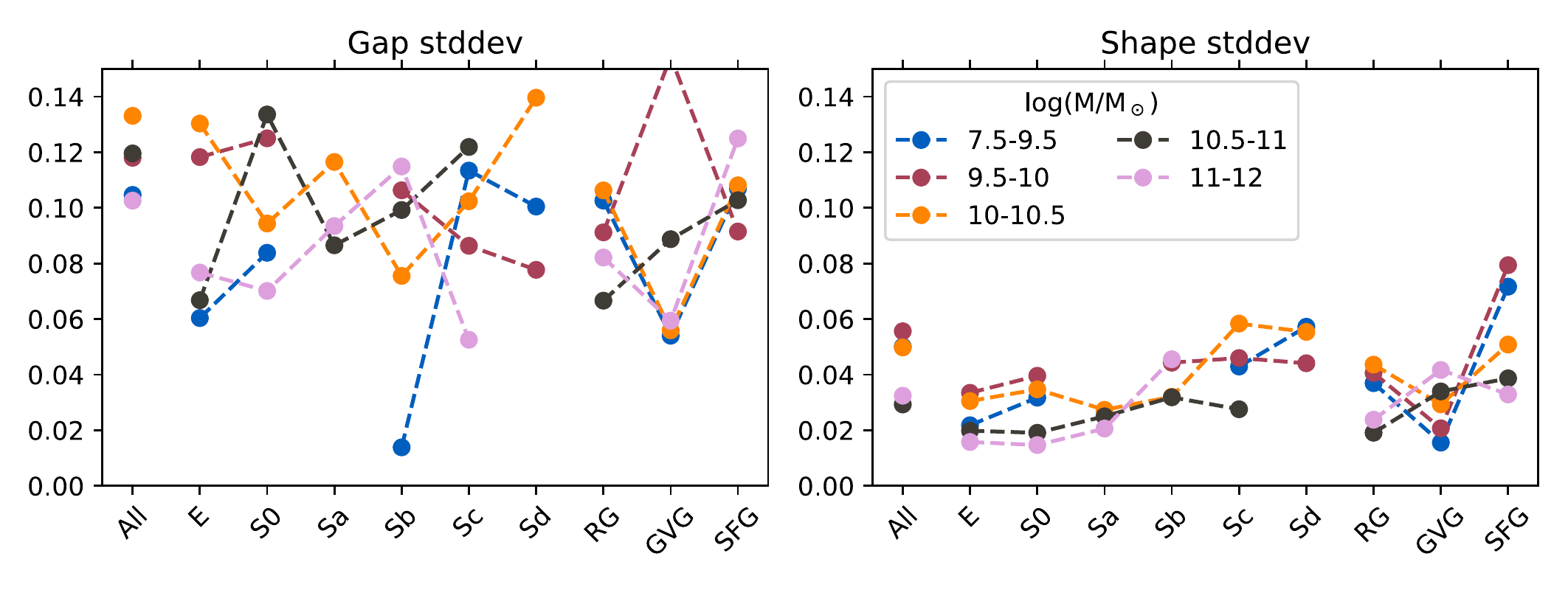}
\caption{Gap variance (left) and shape variance (right) for galaxies in the morphology and mass bins used in this study. The gap variance is the variance between curves due only to the differences on their average values, whereas the shape variance is due to the shape of the ChEH.
}
\label{fig:variance}
\end{figure*}

In Figure \ref{fig:variance} we show the values of these two variances for the mass, morphology and SFS bins used in this paper. We can observe that gap variances are generally higher than shape variances and they also contain more variance within their values. They also do not appear to exhibit any clear trends with morphology, mass or SFS.

The shape variance, on the other hand, does exhibit interesting trends. There is a trend for higher mass bins to have lower shape variance, showing that more massive galaxies have more similar ChEHs. There is also a trend with morphology, where earlier type galaxies have a lower shape variance compared to late types. When comparing between RG, GVG and SFG we see that low mass SFG have a lot more shape variance than the rest, showing also a much more pronounced separation as a result of mass than for RG and GVG.

This suggests, as we find for the mass bins, that later type and star-forming galaxies have a higher diversity in chemical evolution histories but we need to consider one particular caveat of our analysis that could be affecting this result. Earlier type galaxies are generally "dead" galaxies whose metallicity has long stopped evolving. The difference in the shape of their chemical evolution then, if any exists, would have happened at high LBT where our time resolution is much worse.

It is possible, then, that the differences in morphology for the shape variance are an artefact due to the limitations of our analysis.

\section{Discussion}
\subsection{Caveats due the methodology}
The results presented in this paper are strongly model-dependent. SSP fitting can yield different results depending on the fitting code, the templates used to perform the fit and the conditions that we impose such as the IMF or the dust attenuation curve \citep{Walcher2011, Conroy2013, Sanchez2016a, Sanchez2020}.

In addition to these problems, there are intrinsic degeneracies in the observed spectroscopic properties among stellar populations with different physical properties.
One of the most important of these degeneracies is the age-metallicity degeneracy, where a stellar template can produce a similar spectrum to one which is either younger and more metallic or older and less metallic.
Despite these problems, SSP fitting codes are broadly compatible in their results, though care needs to be taken to compare them depending on their different features \citep{Sanchez2016a}.

The factor that has the most impact on the quantitative results, however, is the metallicity range of our templates. The MILES templates are composed from observed stars. This produces more reliable templates to fit as we do not introduce a dependence on the reliability of the stellar evolution and atmosphere models, which is the case with synthetic spectra. However, they are limited to the range and sampling of physical parameters that we have available.
As a result, the set of templates we have chosen from the MILES database have metallicity values no lower than [Z/H] = -0.4 and no higher than 0.2. While this is more than acceptable for fitting the current composition of a galaxy, we are exploring the history of metal enrichment for the galaxies in our sample. We cannot obtain values for the stellar metallicity lower than -0.4 throughout a galaxy's lifetime, which is simply not a realistic value for the early populations even if we discount the initial star formation with pristine gas.

Even so, this caveat  does not completely invalidate the results we are presenting. The absolute values of the metallicity can certainly not be taken as reliable measurements, but the trends that they exhibit should be valid. A stellar population with [Z/H] lower than -0.4 which is present within a spatial element will result in higher weights for the lower metallicity templates for that spatial element. In addition, the age-metallicity degeneracy will most probably shift some weights towards younger stellar populations (to compensate for the lack of lower metallicity populations). This may affect the SFH more than the ChEH. However, as demonstrated by \cite{Ibarra-Medel2019}, based on hydrodynamic simulations, the effect is not dramatically strong.

The sampling in the time dimension covers a wide range of ages, but it is uneven, with much shorter intervals for recent ages.
This is by construction. This sampling was adopted since our ability to distinguish between different stellar populations is less precise for older than for younger stellar populations. Thus, stellar populations differentiate themselves following an almost logarithmic sampling. In other words, two stellar populations of an age of hundreds of Myrs differentiate between themselves well when they have a difference in age of $\sim$100 Myr. For stellar populations of few Gyrs, they present different spectra when they have differences in age of more than $\sim$1 Gyr.

In conclusion, the degeneracies and uncertainties in the method employed here mean that we cannot take quantitative measurements at face value and individual features of a single galaxy's ChEH or MZR (t) curve can well be artefacts. However, the qualitative behaviour of the averaged curves and their features which are consistent between mass and morphology bins should not arise from artefacts related to the fitting method employed. This is a similar conclusion to that of \cite{Ibarra-Medel2019}, based on simulations.
It would certainly be of interest, however, to repeat this study using a different set of SSP and/or a different galaxy sample such as MaNGA, SAMI or AMUSING++ as future work on this topic.
We will perform this exploration in further studies.

\subsection{Previous explorations of MZR(t)}
Existing studies on how the stellar MZR changes through time are \cite{Panter2008} and \cite{Vale-Asari2009}, for a similar analysis to ours, and VANDELS \citep{Cullen2019}, which fit stellar templates to galaxies observed at high redshift. Our results are broadly compatible with these studies from a qualitative point of view. 

All studies, including the current one, agree on that at high redshifts the MZR was steeper than now. They agree too in that massive galaxies were enriched faster than lower-mass ones.
However, \cite{Vale-Asari2009} does not see the flattening of the MZR for low-mass galaxies. This is expected as their sample only includes galaxies down to $10^{10}$ M$_\odot$. This is well above the value at which the MZR appears to flatten in our results. On the other hand, VANDELS does appear to show some flattening at low masses, starting from around $10^{9.5}$ to $10^{10}$ M$_\odot$. This is consistent with our results.

There are large discrepancies as to the quantitative values, with our results having a narrower range of metallicities and especially our metallicity values for high redshifts are much higher, not going below -0.4 as compared to VANDELS or \cite{Vale-Asari2009} which reach values of -1.0 for high redshifts. This discrepancy is expected as a result of the choice in stellar templates as described above, as our SSP are composed from observed stars which limits our range of metallicities to between -0.4 and 0.2. All the above mentioned studies use synthetic templates which provide as wide a range of physical parameters in exchange for a dependence on theoretical models.

\subsection{[Z/H] (t) convergence in SFGs along the last few Gyr}

One of the most interesting results we report in this study refers to the metallicity evolution for star-forming galaxies compared to those that are retired or populate the green valley. These show a convergence in their values at recent times, with the more massive galaxies reaching a maximum early and the less massive ones only just becoming enriched to that value. The GVG and RG, on the other hand, show a more stratified evolution, especially the retired galaxies. What makes this result particularly relevant is the fact that this behaviour is observed only when we separate galaxies by star-forming status.
By comparing Figs. \ref{fig:zh_morph_all} and \ref{fig:zh_sf} we can see that there is a marked difference in the metallicity values for the most massive bin of SFG and the values for Sa-b galaxies. Since all SFG of that mass are in the aforementioned mass-morphology bins it means that the difference, which produces the convergence of values with the less massive bins, is due only to their star-forming status.

Another effect that is observed is a slight, steady, decrease in metallicity for the more massive bins after the initial enrichment. This effect can be seen clearly for the SFG and for the $10^{10.5-11}$ M$_\odot$ bin for the GVG. The GVG for three bins also appear to show some convergence to a lesser extent than the SFG. Given that the GVG are generally considered to be transitioning from the SFG to the RG we can consider this behaviour to be tied to that of the SFG.

The convergence observed suggests that there is some regulatory mechanism that limits how high a metallicity a star-forming galaxy can achieve. Looking at the Sc galaxies in Fig. \ref{fig:zh_morph_all} we might be tempted to claim that the convergence effect is nothing more than the result of mixing morphologies, as the more massive galaxies have a flatter evolution and the less massive galaxies are still growing, but this interpretation does not explain the convergence nor why the massive Sa-b galaxies drop about 0.1-0.2 dex in metallicity when considering only the star-forming galaxies, a drop that produces the observed convergence. It is more likely, then, that Sc show the same behaviour because they are dominated by SFG.

It has been shown that stellar feedback in the form of metal-rich outflows removes metals from galaxies which can result in a regulatory process where the metals can achieve an equilibrium between production and removal. This has been proposed as a mechanism that explains or contributes to the flattening of the nebular MZR at high galaxy masses \citep{Tremonti2004,Mannucci2010,Sanchez2013,Belfiore2017,Barrera-Ballesteros2018}.
While this mechanism undoubtedly contributes to the convergence it is not enough by itself. The reason for this is the differences between SFG-GVG-RG, with the SFG collapsing to a value of metallicity much lower than for GVG and especially RG. In order to produce the observed results a mechanism unique to galaxies with a sustained star-formation is required. The two most massive bins of RG clearly collapse to a maximum value, an effect that can be attributed to the regulation via only outflows and quenching.

As for the additional mechanism for SFG, we believe the easiest answer would be inflows of pristine gas which sustains star-formation. This mechanism has long been proposed to explain how many galaxies appear to maintain a steady star formation rate over time periods much longer than the time it would take to exhaust the available gas within them as well as to reduce the metallicity reached by the formed stars.
These observations have motivated the development of chemical evolution models generally referred to as ‘gas regulatory’ or ‘bathtub’ models  \citep{Finlator2008, Lilly2013, Dekel2013, Peng2014, Barrera-Ballesteros2018, Belfiore2019, ZaragozaCardiel2019, ZaragozaCardiel2020}
Within these models there is a sustained inflow of pristine gas from the halo into the galaxy which provides the material required to maintain star formation over long times, as well as outflows which regulate the gas content.

This scenario does conform to what we observe in our results. As the original gas in the disc is used up to form stars it gets progressively more metallic from the previously formed stars polluting their environment. Once the original gas starts to be exhausted the inflow mechanisms start to become more relevant to sustain further star-formation and newer stars no longer necessarily have higher metallicities than the current population. This can produce the observed convergence, as well as the subtler steady decrease in metallicity after the initial enrichment.

The fact that the decrease is only observed for the more massive bins is counter-intuitive as more massive galaxies have more massive haloes. Thus, these mechanisms should be less effective. One possible explanation is that the less massive galaxies have a higher fraction of the primordial gas in their disk. As a result, inflow-sustained star formation does not diminish the metallicity because these galaxies have not yet reached their maximum values as a result of their delayed evolution. In this manner, even though they also have inflows, this does not produce the decrease in metallicity that is observed for the more massive galaxies which exhausted their primordial gas reserve long ago.

Another explanation is that the observed effect is a result of high mass SFG being a peculiar sample. Downsizing predicts that the more massive galaxies should form stars faster, which combined with less efficient gas inflows means that they should exhaust their gas very quickly. In order for a galaxy to maintain a high SFR while being in the most massive bin it would require an unusually massive inflow sustained through its lifetime. In this scenario, the "normal" galaxies do not populate this particular bin at all, leaving only the outliers. The behaviour of these galaxies would therefore not be representative of galaxies in general, avoiding the discrepancy.

A potential caveat to this interpretation, which in turn affects all results derived from the SFG-GVG-RG separation, is that we are analysing the history of a galaxy based on current and transient properties. We evaluate as star-forming those galaxies that have a high value of EW$_{\mathrm{H\alpha}}$ (above 6 \AA), but the equivalent width traces only ongoing star formation. However, recent results have shown that indeed most current SFGs have been SF throughout their life time (e.g. \citealt{Pandya2017, Sanchez2019} and references therein).
This ensures that the results we present obtained by tracing the evolution of a bin of currently star-forming galaxies are valid, as statistically most of these galaxies have belonged to this bin throughout their history.

\subsection{Caveats and effects of the morphology}

A similar caveat to the segregation by the star-forming status can be found for the separation by morphology. The morphology classification is determined based on the current appearance that the galaxies in our sample exhibit. However, fossil record techniques do not provide information to infer the morphology of a galaxy in the past at least with the spatial resolution and binning needed to recover the resolved SFH. Fossil record data obtained from high resolution IFS observations such as MUSE might be able to achieve this, though it would still be limited by the effect that radial migrations could have on the distribution of stars.

Understanding the morphological evolution of galaxies is an ongoing effort in the field. \cite{Conselice2014} reviews the topic, including surveys of multi-redshift observations using deep-field imaging to determine the morphologies of galaxies in the past, such as CANDELS \citep{Grogin2011, Koekemoer2011}. The main results are that the current Hubble type classification stops being representative of galaxies at around redshift 2, with the number of peculiar galaxies growing until it dominates the population. As we move towards closer times the fraction of peculiar galaxies drops drastically while that of disk galaxies and especially spheroidals grow. This suggests that the bulk of morphology evolution is not a process of galaxies "switching" between the morphological types we currently observe, but an evolution towards them.
When we show the evolution of the metallicity or the MZR within a morphological bin, then, it should be understood as the evolution of galaxies whose shape ends up as that morphological type.

\subsection{Relation between the ChEH and the SFR'}

The derivative of the mass (SFR') is somewhat equivalent to the SFH of a galaxy, as rapid changes in the mass of a galaxy are necessarily associated to episodes of star formation. The difference is that while the SFH only traces the addition of new stars the SFR' also includes the death of preexisting stars. The latter phenomenon, however, manifests as a steady decrease in M$_*$ as the lifetimes of stars depend on their mass. Because of this if a peak appears in the SFR' it represents a burst of star formation.

The peaks in SFR' that we observe appear to be associated with a jump between two somewhat flat regimes of the metallicity growth. All galaxies may have individual bursts of star formation throughout their life-times. However, this appears to be a global burst of star-formation which stands out even after averaging different galaxies with different histories. The same conclusion can be reached for the jump in metallicity: for such a feature to consistently appear when we average over several galaxies indicates that the phenomenon is global. This does not mean that all galaxies should exhibit these two features individually. For the effect to arise on the average it would be enough if it was more likely for galaxies to have a burst of star formation and a decrease in metallicity at these times than others.


This phenomenon correlates with the morphological type and especially the mass of the galaxies. It is more prominent for later and less massive galaxies. Indeed, the more massive bins do not exhibit a visible jump in metallicity except for Sc galaxies. The higher prominence for late type and low-mass galaxies suggests that this may be related to the gas content of galaxies, as both of these properties correlate with higher gas content. Even if this is the case, however, there is no clear link between a galaxy having a high gas content and it suffering an episode of star formation. This is especially so for a galaxy with a stellar metallicity which greatly differs from the value that would be expected from its previous evolution. A massive inflow of pristine, or very poor metal gas could cause an important SFR burst and, consequently, a fast ($\sim$ Myr) increase of the metallicity of the ISM. Therefore, most stars formed during the burst could present high metallicity. If the metallicity of the ISM before the burst is low (as is the case of low-mass galaxies), the metallicity increase is higher. Moreover, if well-mixed outflows occur during the burst, the metallicity increase is even higher.

\section{Summary and conclusions}

\begin{itemize}
\item We have used fossil record techniques to study the chemical evolution of a sample of galaxies observed with IFS (CALIFA). The analysis allows us to measure the stellar metallicity at different cosmological times.

\item The chemical enrichment history of galaxies shows a faster evolution for the more massive and earlier type galaxies, which also have higher terminal values of the metallicity. The lower mass and latter type galaxies have lower metallicities and appear to still be steadily increasing their metallicity, a sign that they are still assembling their mass.

\item Star-forming galaxies appear to have an upper limit of metallicity towards which they converge, regardless of stellar mass. The most massive bins show a noticeable decrease in metallicity for SFG, as well as for GVG to a lesser degree.

\item The MZR shows a slope change resulting from the more massive galaxies evolving faster than the less massive ones. Upon separating the galaxies into morphology bins, however, this behaviour becomes much less pronounced, suggesting that morphology is an important factor in determining how fast galaxies evolve.

\item Galaxies evolve more similarly in the outskirts compared to the centre or the effective radius, with low mass galaxies showing a positive metallicity gradient.

\item The chemical enrichment histories of galaxies differ more in terms of absolute value rather than in their shapes. While their differences due to their absolute values have no correlation to mass, morphology, or star-forming status there is a clear correlation regarding their difference in shape. Earlier, more massive and retired galaxies have more similar evolutions, while later, less massive and star-forming galaxies have more differences between the shapes of their enrichment curves.
\end{itemize}

\section*{Acknowledgements}

We are grateful for the support of a CONACYT grant CB-285080 and FC-2016-01-1916, and funding from the PAPIIT-DGAPA-IN100519 (UNAM) project.
L.G. was funded by the European Union's Horizon 2020 research and innovation programme under the Marie Sk\l{}odowska-Curie grant agreement No. 839090. This work has been partially supported by the Spanish grant PGC2018-095317-B-C21 within the European Funds for Regional Development (FEDER).
We thank P. Tissera, N. Vale-Asari and J. E. Beckman for their many helpful comments and discussion on the content of the paper.

\section*{Data Availability}
The data used in this article consists of the publicly available DR3 of the CALIFA survey as well as extended surveys such as PISCO. CALIFA DR3 data cubes can be accessed at  \url{https://califaserv.caha.es/CALIFA_WEB/public_html/?q=content/califa-3rd-data-release}, with direct access to the V500 version used in this article available at \url{ftp://ftp.caha.es/CALIFA/reduced/V500/reduced_v2.2/}.

The extended survey data can be shared on reasonable request by contacting SFS and LG.

     \bibliographystyle{mn2e}

\label{lastpage}

\end{document}